\documentclass{article}

\usepackage{arxiv}

\RequirePackage[OT1]{fontenc}
\RequirePackage{amsthm,amsmath}
\RequirePackage[numbers]{natbib}
\RequirePackage[colorlinks,citecolor=blue,urlcolor=blue]{hyperref}
\usepackage{amsmath}
\usepackage{amsbsy}
\usepackage[noend]{algpseudocode}
\usepackage{graphicx}
\usepackage{mathtools}
\usepackage{amssymb}
\usepackage{eqnarray}
\usepackage{listings}
\usepackage{units}
\usepackage{footnote}
\usepackage{blkarray}
\usepackage{comment}
\usepackage{graphicx,amsmath,amssymb}
\usepackage{algorithm}
\usepackage{longtable}
\usepackage{subcaption}

\newcommand{\E}{\ensuremath{\mathbb{E}}}
\newcommand{\R}{\ensuremath{\mathbb{R}}}

\newcommand{\mc}[1]{\ensuremath{\mathcal{#1}}}

\newcommand{\kl}[2]{\ensuremath{D(#1 \mid \mid #2)}}

\newcommand{\mcal}[1]{\ensuremath{\mathcal{#1}}}

\newcommand{\ra}{\rightarrow}

\newcommand{\sm}{\setminus}

\newcommand\independent{\protect\mathpalette{\protect\independenT}{\perp}}
\def\independenT#1#2{\mathrel{\rlap{$#1#2$}\mkern2mu{#1#2}}}

\newtheorem{Theorem}{Theorem}

\newtheorem{Example}{Example}[section]

\newtheorem{Definition}{Definition}

\title{Direct and Indirect Effects -- An Information Theoretic Perspective}

\author{
  Gabriel Schamberg\\
  Picower Institute for Memory and Learning\\
  Massachusetts Institute of Technology\\
  \texttt{gabes@mit.edu}
   \And
  William Chapman\\
  Scripps Institution of Oceanography\\
  \texttt{wchapman@ucsd.edu}
   \And
  Shang-Ping Xie\\
  Scripps Institution of Oceanography\\
  \texttt{sxie@ucsd.edu}
   \And
  Todd P.~Coleman \\
  Department of Bioengineering\\
  University of California, San Diego\\
  \texttt{tpcoleman@ucsd.edu}
}

\begin{document}
\maketitle

\begin{abstract}
Information theoretic (IT) approaches to quantifying causal influences have experienced some popularity in the literature, in both theoretical and applied (e.g. neuroscience and climate science) domains. While these causal measures are desirable in that they are model agnostic and can capture non-linear interactions, they are fundamentally different from common statistical notions of causal influence in that they (1) compare distributions over the effect rather than values of the effect and (2) are defined with respect to random variables representing a cause rather than specific values of a cause. We here present IT measures of direct, indirect, and total causal effects. The proposed measures are unlike existing IT techniques in that they enable measuring causal effects that are defined with respect to specific values of a cause while still offering the flexibility and general applicability of IT techniques. We provide an identifiability result and demonstrate application of the proposed measures in estimating the causal effect of the El Ni\~no-Southern Oscillation on temperature anomalies in the North American Pacific Northwest.
\end{abstract}



\section{Introduction}
Consider a directed acyclic graph (DAG), where nodes represent random variables and edges represent a direct causal influence between two variables. We here discuss the problem of \emph{quantifying} these causal influences. This problem has received considerable attention in a variety of communities; for the sake of exposition, we coarsely categorize methods as either statistical (i.e. those summarized by \cite{pearl2009summary}) or information theoretic (IT) (i.e. those measured in units of bits or nats) \cite{janzing2013quantifying,ay2008information,pocheville2017comparing}. When viewed from an applications perspective, these two approaches are quite different. Statistical approaches are common in epidemiology and economics \cite{hernan2010causal,imbens2009recent}, whereas IT methods appear in the study of complex natural systems, for example climate scientic \cite{hlavavckova2007causality,runge2015quantifying} or neuroscientific \cite{kim2014dynamic,wibral2014directed}. The fundamental difference in perspectives that gives rise to this disparity is not well presented in the development of IT methodologies.

To illustrate this difference, consider a simple example with a two node graph $X\ra Y$, where $X\in\{0,1\}$ represents whether or not an individual has won the lottery and $Y \in \mathbb \R$ represents that individual's average monthly spending (assume for clarity that there are no confounding factors, i.e. observing a lottery winner is equivalent producing a lottery winner by means of intervention). A statistical measure such as the average causal effect (ACE) \cite{holland1988causal,pearl2009causality} would seek to answer the question ``What is the effect of \emph{winning} the lottery on spending?'' by comparing the average spending of lottery winners ($X=1$) against the average spending of lottery non-winners ($X=0$): $\E[Y \mid X =1]-\E[Y \mid X =0]$. We would of course expect this to be quite large. It is important to note that the ACE is defined irrespective of the marginal distribution of $X$, meaning that the probability with which $x$ occurs has no bearing on the effect of $x$ on $Y$. An IT approach addresses a subtly different question: ``What is the effect of the lottery on spending?'' In other words, an IT measure considers the effect of the random variable representing whether or not one wins the lottery on spending. Specifically, the effect of $X$ on $Y$ would be given by the mutual information (MI), $I(X;Y)$ (see \cite[Sec. 2, {\bf (P2)}]{janzing2013quantifying}). Using a simple IT inequality, we get that the MI is bounded above the Shannon entropy, $H(X)$. Given that the odds of winning the lottery are essentially a point mass, which has zero Shannon entropy, we have $I(X;Y)\le H(X) \approx 0$. In words, because so few people win the lottery, an IT measure indicates that the lottery has a negligible effect on spending. In this regard, statistical measures consider the effect of a \emph{specific cause}, whereas IT measures have historically considered the effect at a \emph{systemic level}.

A second difference is that, whereas statistical approaches typically measure causal effects on the \emph{value} of an outcome, IT approaches measure the causal effect on the \emph{distribution} of an outcome. Each of these approaches comes with benefits and drawbacks. With statistical approaches, the units are preserved (in the previous example, the units of the ACE are dollars). While IT measures yield the less interpretable unit of bits, they are able to capture more complex causal effects, for instance the effect that a variable has on the variance of another. Acknowledging this difference helps to understand the disparity between the applications of statistical and IT measures. When evaluating the causal link between smoking and cancer, the number of bits of information shared by the smoking and cancer variables may not be as useful as knowing the extent to which quitting smoking decreases the likelihood of cancer. However, when studying the nature of complex natural networks, it may be desirable to use a measure that can capture higher order causal effects.

Of course, not all causal inference approaches fall neatly into this coarse categorization. There has been considerable work in the statistics literature on distributional effects, most commonly using quantile effects \cite{chernozhukov2005iv,firpo2007efficient}. Quantile effects measure the difference between a particular quantile of two distributions (for example the median), and thus, like the average causal effect, do not capture the effect that an intervention may have on a distribution as a whole. While approaches measuring the $L_1$ distance between distributions (given by the integral/sum of absolute differences) \cite{kim2018causal} and the difference in Gini indices of distributions \cite{rothe2010nonparametric} provide reasonable alternatives to the proposed approach, they offer no insight into the nature of information theoretic measures of causal influence in the broader context of causal inference. It should also be noted that not all statistical measures of causal effect rely on the specification of two values of a cause. Studies using stochastic intervention effects allow for interventions to affect the \emph{distribution} of a cause \cite{munoz2012population}.

The problem considered presently is distinct from that addressed by popular time series analyses such as Granger causality \cite{granger1969investigating}, directed information \cite{marko1966theorie,marko1973bidirectional,massey1990causality}, and transfer entropy \cite{schreiber2000measuring}. Rather than evaluating the effects of interventions in a causal model, these methods rely on time-lagged correlations or mutual informations. While scenarios exist where these methods coincide with approaches based on interventions, they are not equivalent in general. In this paper we focus on methods based on interventions and refer the reader to \cite{lizier2010differentiating,eichler2010granger,schamberg2019measuring} for further discussion on the relationship between the interventional and non-interventional approaches.

In the present work we seek to endow IT measures with the ability to measure \emph{specific} causal effects. Furthermore, we show that existing IT measures of causal influences are ill-equipped for distinguishing direct and indirect effects. Following a parallel storyline to that of Pearl \cite{pearl2001direct}, we provide measures of the total, (natural and controlled) direct, and natural indirect effects. We show that these measures do not fundamentally change the underlying IT perspective on causality, but enable obtaining ``higher resolution'' measures of causal influence. In doing so, we provide increased clarity to the aforementioned differences between IT and statistical causal measures. We showcase how the framework can be used in practical contexts, focusing on the evaluation of the causal effect of the El Ni\~no-Southern Oscillation (ENSO) on land surface temperature anomalies in the North American Pacific Northwest (PNW). Our results confirm the scientific consensus that both ENSO phases affect PNW land surface temperatures asymmetrically. Furthermore, using a conditional version of the proposed measures, we show the presence of a ``persistence signal'' across two-week average temperature anomalies that is modulated by the El Ni\~no phase. This result both demonstrates the value of the proposed framework and provides direction for future studies focused on climate scientific findings.

The remainder of the paper is structured as follows: Section \ref{preliminaries} introduces notation and provides background on the relevant works on the quantification of causal effects. Section \ref{sec:defs} provides definitions of novel measures of causal influence along with a number of relevant properties and extensions. Section \ref{examples} presents intuitive examples demonstrating the utility of the proposed perspective. Section \ref{sec:climate} presents our case study applying the proposed measures to measure the effect of ENSO on PNW temperature anomalies. Finally, Section \ref{sec:conclusion} contains concluding remarks.

\section{Preliminaries}\label{preliminaries}
\subsection{Notation and Problem Setup}\label{notation}

\begin{figure}
  \begin{center}
    \includegraphics[width=0.35\linewidth]{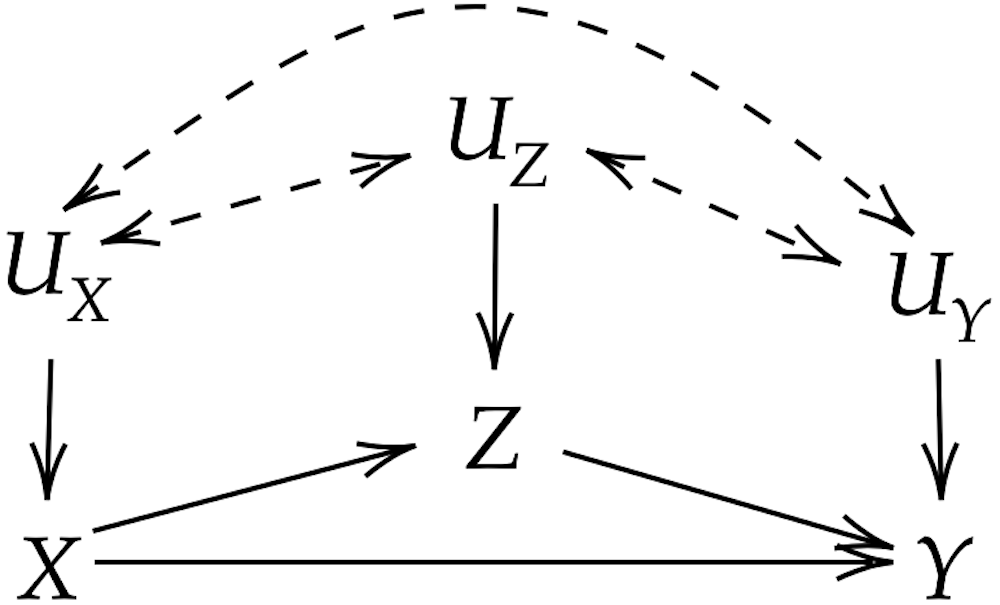}
  \end{center}
  \caption{DAG \mc{G} representing a mediation model}\label{fig:dag}
\end{figure}

We will be developing techniques for measuring the causal influence of $X\in \mcal{X}$ upon $Y\in\mcal{Y}$ in the presence of a mediating variable $Z\in \mcal{Z}$ using the DAG $\mcal{G}$ depicted in Figure \ref{fig:dag}. Without loss of generality, $Z$ may represent a collection $(Z_1,Z_2,\dots,Z_k)\in \mcal{Z}_1\times\mcal{Z}_2\times\dots\times\mcal{Z}_k=\mcal{Z}$ of all mediating variables. Define the parent sets of $X$, $Z$, and $Y$ as $PA_X = U_X$, $PA_Z=\{X\}\cup U_Z$, and $PA_Y=\{X,Z\}\cup U_Y$. Dashed double headed arrows in Figure \ref{fig:dag} are used to indicate unknown dependencies between $U_X\in\mcal{U}_X$, $U_Y\in\mcal{U}_Y$, and $U_Z\in\mcal{U}_Z$ (including the possibility of $U_S \cap U_T\ne \emptyset$ for $S,T\in\{X,Y,Z\}$). We may use the shorthand $U=U_X\cup U_Y\cup U_Z\in\mcal{U}$. For simplicity, we assume that all variables are discrete with arbitrary finite supports. When appropriate probability densities exist, extending the proposed definitions to continuous or mixed random variables is straightforward. Extending estimators of the proposed measures to the continuous case is less straightforward and is not presently considered. In general, let $p$ be the probability mass function (pmf) for all variables in the graph (i.e. $X,Y,Z,U\sim p$), capital letters represent random variables, and lowercase letters represent their realizations. For example, $p(x\mid pa_X)$ gives the conditional probability of the event $X=x$ given that its parents took on values $pa_X$. We further assume that $p$ satisfies the causal Markov condition with respect to \mc{G} \cite{pearl2009causality}, with $p(x,y,z,u) = p(u)p(x\mid u_X)p(z\mid x,u_Z)p(y\mid x,z,u_Y)$. We use a hat to indicate the $do$-operator, which represents taking the action of forcing a variable to assume a particular value by means of intervention. For example, $p(y\mid \hat{z})=p(y\mid do(Z=z))$ gives the probability of $y$ given that $Z$ is forced to take the value $z$, irrespective of the probability with which that value occurs. When working with distributions utilizing the $do$-operator, a set of rules known as the $do$-calculus can be used to identify if and how the interventional distributions correspond to observational distributions that do not utilize the interventions. While the reader is referred to \cite[Sec. 3.4]{pearl2009causality} for the complete $do$-calculus, we provide a description of the rule which enables swapping interventions for observations in Appendix \ref{app:docalc}.

The entropy of a random variable $Y$ and conditional entropy of $Y$ given $X$ are given respectively by $H(Y)=-\sum_{y}p(y)\log p(y)$ and  $H(Y\mid X)=-\sum_{x,y}p(x,y)\log p(y\mid x)$. It is worth noting that the conditional entropy yields the \emph{expected} uncertainty of $Y$ given $X$, and is not to be confused with $H(Y\mid X=x)=-\sum_{y}p(y)\log p(y\mid x)$, which gives the uncertainty of $Y$ when conditioning on a \emph{particular value} of $X$. For two distributions $p$ and $q$ over $\mcal{Y}$, the KL-divergence (also known as relative entropy) from $p$ to $q$ represents the excess number of bits needed to represent $Y$ if the distribution is assumed to be $q$ when it is in fact $p$, and is given by $D(p(Y)\mid\mid q(Y))=\sum_y p(y)\log \nicefrac{p(y)}{q(y)}$ \cite{cover2012elements,mackay2003information}. The KL-divergence is zero if and only if $p(y)=q(y)$ for all $y$ for which $p(y)>0$, and is deemed infinite if there exists a $y$ such that $p(y)>0$ and $q(y)=0$. We use $Bern(\alpha)$ to represent the distribution of a Bernoulli random variable with parameter $0<\alpha <1$. For the KL divergence between two Bernoulli random variables with parameters $\alpha$ and $\beta$, we will use the shorthand $D(\alpha\mid\mid\beta)$. Finally, the mutual information (MI) between $X$ and $Y$ is given by $I(X;Y)=H(Y)-H(Y\mid X) = \sum_x D(p(Y\mid X=x)\mid\mid p(Y))p(x)$. These equivalent definitions of MI give rise to two interpretations: (i) the average reduction in uncertainty in $Y$ obtained by conditioning on $X$ and (ii) the average increased ability to predict $Y$ resulting from conditioning on $X$. It is worth noting that (barring some technical details), these definitions can be applied to continuous valued random variables by substituting integrals for sums and probability density functions for pmfs.

\subsection{Direct and Indirect Effects}\label{sec:pearl}
Building upon the work of Robins and Greenland \cite{robins1992identifiability}, Pearl \cite{pearl2001direct} formalized definitions of direct and indirect effects in the context of graphical models. Such a distinction is useful in disentangling the mechanisms via which causal influences occur. A canonical example is presented by Hesslow \cite{hesslow1976two}, wherein a birth control pill is suspected of directly increasing the likelihood of thrombosis in women, while simultaneously reducing thrombosis through its prevention of pregnancy (which is positively linked to thrombosis). In each of Pearl's definitions, the magnitude of the causal effect is specified for a specific value $x$ and is measured with respect to a reference (or baseline) value $x^*$. The simplest of these measures is the total effect (TE) of $X=x$ on $Y$ given by $\E[Y\mid \hat{x}] - \E[Y\mid \hat{x}^*]$. The TE yields the answer to a very concise causal question, namely ``How much would we expect the value of $Y$ to change if we were to change $X$ from $x^*$ to $x$?'' As indicated by the name, the TE does not distinguish effects that $x$ has on $Y$ directly from those that occur via a mediating variable $Z$. As such, Pearl proceeds to define the controlled direct effect (CDE) of $x$ on $Y$ with mediator $z$ as $\E[Y\mid \hat{x}^*,\hat{z}]-\E[Y\mid \hat{x},\hat{z}]$. Once again, this measure addresses a clear causal question: ``How much would we expect the value of $Y$ to change if we were to change $X$ from $x^*$ to $x$, but kept $Z$ at a fixed value $z$?'' While this is an intuitive notion of direct effect, it is important to note that it requires the intervention $do(Z=z)$. Given that it may be of interest to know the direct effect that occurs when the mediating variable is \emph{not} controlled for, Pearl defines the \emph{natural} direct effect (NDE) as $\E[Y\mid \hat{x},Z_{x^*}]-\E[Y\mid \hat{x}^*]$, where $Z_{x^*}$ is the value $Z$ would have taken had $X$ been $x^*$. Using this notion of simultaneously assigning a value to $X$ and allowing $Z$ to take the value it would under a different $X$, Pearl defines the natural indirect effect (NIE) as $\E[Y\mid \hat{x}^*,Z_{x}]-\E[Y\mid \hat{x}^*]$. In words, the natural indirect effect represents the expected change in $Y$ resulting from changing $Z$ from the value it would take under $x^*$ to the value it takes under $x$ while leaving $X$ fixed at $x^*$. 


\subsection{Information Theoretic Notions of Causal Influence}
While there is a considerable body of work developing IT techniques for measuring causal influence, we here focus on information flow \cite{ay2008information} and causal strength \cite{janzing2013quantifying}.

\subsubsection{Information Flow}
Drawing on the relationship between mutual information and statistical dependence, Ay and Polani \cite{ay2008information} define an IT notion of \emph{causal independence}, which unlike mutual information, is directed. Their definitions rely heavily on the post-interventional distribution, which dictates a truncated factorization of a joint distribution in the presence of interventions. The information flow (IF) from $X$ to $Y$ is defined as:
\begin{equation}\label{eq:IF}
I(X\ra Y)\triangleq \sum_{x}p(x)\sum_y p(y\mid \hat{x})\log \frac{p(y\mid \hat{x})}{\sum_{x'}p(x')p(y\mid \hat{x}')}
\end{equation}
If all the hats are removed from the above equation, then the standard mutual information is recovered. By using these post-interventional distributions, however, all ``upstream'' dependencies of $X$ are ignored, and thus any relationship between $X$ and $Y$ resulting from confounding variables is removed. Ay and Polani also define a conditional version of IF. Using the mediation model in Figure \ref{fig:dag}, let $V$ be some subset of remaining variables in the graph, i.e. $V\subseteq U\cup\{Z\}$. The IF from $X$ to $Y$ imposing $V$ is given by:
\begin{equation}\label{eq:condIF}
I(X\ra Y\mid \hat{V})\triangleq \sum_v p(v) \sum_{x}p(x\mid \hat{v})\sum_y p(y\mid \hat{x},\hat{v})\log \frac{p(y\mid \hat{x},\hat{v})}{\sum_{x'}p(x'\mid\hat{v})p(y\mid \hat{x}',\hat{v})}
\end{equation}
Noting that $V$ always appears as an intervention, the conditional IF can be interpreted as representing the IF from $X$ to $Y$ when the value of $V$ is controlled. The IF can be extended to measure the flow to and from \emph{sets} of nodes, though at present we only consider the flow from $X$ to $Y$. IF is not to be confused with Marko and Massey's directed information \cite{marko1966theorie,marko1973bidirectional,massey1990causality} or Schreiber's transfer entropy \cite{schreiber2000measuring}, as these do not employ any notion of intervention and are only used in the context of time series.

Within the IF framework, we can treat $I(X\ra Y)$ as a measure of the total effect of $X$ on $Y$ and $I(X\ra Y\mid \hat{Z})$ as a measure of controlled direct effect. While these measures are intuitively analogous to the measures in \cite{pearl2001direct}, it is difficult to formalize the nature of this analogy because we cannot formulate IF measures as the answer to a concise causal question similar to those of the previous section. Furthermore, because the conditional version of IF represents \emph{controlling} a set of variables, IF offers no way to measure the \emph{natural} direct and indirect effects proposed by Pearl.

\subsubsection{Causal Strength}
The causal strength (CS) measure proposed by Janzing et al. \cite{janzing2013quantifying} takes a slightly different approach in that it measures the strength of specific edges in a DAG. We call this an ``edge-centric'' perspective, in contrast with the ``node-centric'' perspective used by IF. To motivate the definition of CS, the authors propose a collection of five postulates that they argue ought to be satisfied by measures of CS. Janzing et al. acknowledge that their postulates need not apply to all reasonable measures of causal influence; as such, any present criticisms of CS can be attributed to differences in the problem formulation. The postulates are briefly summarized here, and the reader is referred to \cite{janzing2013quantifying} for more thorough definitions: {\bf (P0)} If the CS of an arrow is zero, then that arrow may be removed from the DAG without breaking the causal Markov condition. {\bf (P1)} If the entire DAG is given by $X\ra Y$, then the CS is $I(X;Y)$. {\bf (P2)} The strength of an arrow $X\ra Y$ should be defined locally, i.e. it should depend only upon the distributions $p(y\mid pa_Y)$ and $p(pa_Y)$. {\bf (P3)} The CS of an arrow $X\ra Y$ should be lower bounded by the conditional mutual information $I(X;Y\mid PA_Y\sm \{X\})$. {\bf (P4)} If the CS of a set of edges is zero, then the CS of all subsets of those edges should be zero.

Janzing et al. \cite{janzing2013quantifying} proceed to propose a measure of CS that satisfies these postulates. Central to their CS measure is the post-cutting distribution. Formally, let $V=\{V_1,\dots,V_n\}$ be the nodes in a graph, $PA_j^S$ be the subset of parents of $V_j$ for which $V_i\ra V_j \in S$, and $PA_j^{\bar{S}}=PA_j \sm PA_j^S$. Then the post-cutting distribution is given by:
\begin{equation}
p_S(v_1,\dots,v_n)=\prod_{j}\left[
\sum_{pa_j^S}p(v_j\mid pa_j^{\bar{S}},pa_j^S)\left(\prod_{v\in pa_j^S} p(v)\right)
\right]
\end{equation}
The post-cutting distribution factorizes much like the joint distribution $p$ -- however, for nodes at the receiving end of an edge in $S$, they are fed the \emph{marginal distribution} of the node at the other end, rather than the actual value of that node. Using the post-cutting distribution, the CS of a set of edges $S$ is then given by $\mathfrak{C}_{S}=D(p\mid\mid p_S)$, and thus provides a measure of how much excess information is needed to accommodate the severed edges.

Consider CS in the context of the mediation model in Figure \ref{fig:dag}, i.e. $D(p(X,Y,Z,U)\mid\mid p_S(X,Y,Z,U))$ for some set of edges $S\subseteq \{X\ra Y, X\ra Z, Z\ra Y\}$. Within the constraints of the CS framework, one might seek to measure the total, direct, and indirect effects as the strength of the edge sets $S_{TE} = \{X\ra Y, X\ra Z, Z\ra Y\}$, $S_{DE} = \{X\ra Y\}$, and $S_{IE} = \{X\ra Z, Z\ra Y\}$, respectively. To see why this is insufficient, consider an extreme case of the birth control pill example above, where the indirect and direct effects of $X$ on $Y$ are perfectly complementary such that when $x_1=$ ``birth control used'' and $x_2=$ ``birth control not used'' we have $p(y\mid \hat{x}_1)=p(y\mid \hat{x}_2)$ for any $y\in\mcal{Y}$. Any reasonable measure of total effect will conclude that no value of $X$ has an effect on $Y$ -- however, note that from postulate {\bf (P4)}, the total effect (as we have defined it in the CS framework) must be non-zero if either the direct or indirect effect is non-zero. A similar example can be constructed for the insufficiency of CS as a measure of indirect effects by having the effect of $X$ on $Z$ be canceled out by the effect of $Z$ on $Y$. Finally, CS is similar to IF in that it does not yield a clear causal question for which it gives the answer. This is perhaps justified by the decision to define a set of formal postulates that are used to link the properties of CS with our intuitions. However, given that causal influences are likely to be measured in order to obtain a better understanding of the system under study, we find it to be of great practical use to pair causal measures with an easily interpretable causal question for which the measure provides an answer. We will now show that this can be achieved by defining a measure of causal effect of specific values of $X$.

\section{Novel Information Theoretic Causal Measures}\label{sec:defs}
The observation that the MI $I(X;Y)$ does not capture how different values of $X$ may contain different amounts of information about $Y$ has been made in a variety of contexts throughout the literature, including experimental design \cite{lindley1956measure,degroot1962uncertainty}, neural stimulus response \cite{deweese1999measure,vu2009information}, information decomposition \cite{williams2010nonnegative,finn2018pointwise}, measuring surprise \cite{itti2006bayesian}, and most recently, distinguishing between information transfer and information copying \cite{kolchinsky2019decomposing}. Central to each of these works is the development of a notion of MI for a \emph{specific value} of $X$, i.e. $I(x;Y)$. There is, however, no inherent $I(x;Y)$ implied by the definition of $I(X;Y)$ -- to see this, we use the notation of \cite{deweese1999measure} and provide two candidate definitions of $I(x;Y)$ based on the two equivalent definitions of $I(X;Y)$:
\begin{align}
I_1(x;Y) &= D(p(Y\mid X=x)\mid\mid p(Y)) \\
I_2(x;Y) &= H(Y\mid X=x) - H(Y)
\end{align}
It is well understood that, in general, $I_1(x;Y)\ne I_2(x;Y)$. This is clear to see by simply noting that, for any joint distribution $X,Y\sim p$, $I_1(x;Y)\ge 0$ for all $x$, whereas it is possible to have $I_2(x;Y)<0$. In words, the knowledge of a specific value of $X$ will only provide us with a more accurate distribution of $Y$ ($I_1\ge 0$), though it is possible for this distribution to have a greater entropy than the marginal distribution ($I_2<0$). We here use $I_1$ as a foundation for establishing value specific measures of causal influence, and, using the terminology of \cite{kolchinsky2019decomposing}, refer to it as the \emph{specific mutual information} (SMI). Building upon this language in the present context, we refer to the quantities measured by the proposed methods as \emph{specific causal effects}. To our knowledge, the use of SMI in the context of quantifying causal influence is novel. As such, we begin with an informal discussion around the use of SMI for the quantification of causal influence in two-node DAGs, followed by a formal definition of various specific causal effects in a mediation model.

\subsection{Specific Mutual Information in Two-Node DAGs}\label{sec:specific_mi}
Consider a DAG $X\ra Y$ with joint distribution over nodes $X,Y \sim p$, and for the sake of exposition, assume there are no confounding variables. In this simple scenario, when considering the effect of $X$ on $Y$, we can freely exchange interventions for observations (assuming we only consider $x$ s.t. $p(x)>0$), and thus the ACE of $x$ with respect to baseline $x^*$ is given by $\E[Y\mid \hat{x}]-\E[Y\mid \hat{x}^*]=\E[Y\mid x]-\E[Y\mid x^*]$. Once again, this addresses the question of how much the value of $Y$ is expected to change as a result of switching from $x^*$ to $x$. With regard to the CS and IF methods discussed above, both would quantify the effect of $X$ on $Y$ as $I(X;Y)$. Consider the SMI $I_1(x;Y)$ as a measure of the specific causal influence of $x$ upon $Y$ and note the following:

{\bf (I)} We have the equivalence $I(X;Y)=\E[I_1(X;Y)]$, where the expectation is taken with respect to $X$. As such, we can think of the specific causal effect as a \emph{random variable}, whose expectation is the mutual information. In doing so, we are able to capture that different values of $X$ may have different \emph{magnitudes} of causal effect on $Y$, with each of those effects occurring with some probability according to $p(x)$. Moreover, this makes clear that the perspective adopted here is consistent with that of other IT measures.

{\bf (II)} $I_1(x;Y)$ is non-negative for all $x\in \mcal{X}$. Whereas a negative ACE has the clear interpretation of $x$ causing a decrease in the expected value of $Y$, we are measuring influences that $x$ has on \emph{the distribution of $Y$}. Given that there is no obvious notion of a (potentially negative) difference between distributions, we utilize a definition that results in all causal effects having positive magnitude. This serves as a partial justification for using $I_1$, rather than $I_2$, as a foundation. 

{\bf (III)} The SMI does not depend on the value of $Y$, standing in contrast with the information of a single event $i(x;y)\triangleq\log\frac{p(y\mid x)}{p(y)}$ introduced by Fano \cite{fano1961transmission} and its variants, referred to as ``local information measures'' \cite{lizier2008local,lizier2014jidt}. Interpreting local information measures as measures of causal influence is challenging given that they are negative when $Y$ takes on values that are unexpected given $X=x$. We adopt the perspective that, while different values of $X$ may have different levels of effect on $Y$, they can only affect the \emph{distribution} of $Y$, with the specific value $y$ occurring randomly according to an appropriate conditional (or interventional) distribution.

{\bf (IV)} The SMI does not require specifying a reference value $x^*$. Instead, we can view SMI as measuring the causal effect of $x$ as compared with the $X$ that would have occurred naturally. This suggests an intuition for the appearance of IT measures of causal influences in complex natural networks -- values of $X$ that are seen as \emph{changing the course of nature} will be assigned a large causal influence. Given that we can (in this setting) exchange observation for intervention, we can view the SMI as comparing the effect of an intervention $\hat{x}$ with a random (i.e. non-atomic) intervention $\hat{X}$ with $X\sim p$ (see \cite{pearl2009causality,peters2017elements} for discussions on random interventions).

{\bf (V)} The SMI addresses a very clear causal question: ``How much different would we expect the distribution of $Y$ to be if, instead of forcing $X$ to take the value $x$, we let $X$ take on a value naturally?'' Stated more compactly: ``How much would we expect performing the intervention $do(X=x)$ to change the course of nature for $Y$?''

{\bf (VI)} We can interpret the SMI as comparing a ground truth distribution of $Y$ conditioned on $x$ ($p(Y\mid x)$) with a counterfactual distribution wherein nature was allowed to run its course ($p(Y)$). This works well with the interpretation of the KL-divergence as a measure of excess bits resulting from encoding $Y$ using the distribution that is not the true distribution from which $Y$ is sampled. The use of the KL-divergence is further justified in this context by the fact that the logarithmic loss is unique in its ability to capture the benefit of conditioning on $X$ in the prediction of $Y$ \cite{jiao2015justification}.

{\bf (VII)} Finally, we note that $I_1(x;Y)=0$ if and only if $p(y\mid x)=p(y)$ for all $y$ for which $p(y)>0$. By contrast, it is possible to have $I_2(x;Y)=0$ and $p(y\mid x)\ne p(y)$. The following example illustrates why this is undesirable:
\begin{Example}
Consider a two-node DAG $X\ra Y$ with $X\sim Bern(\nicefrac{1}{7})$, $Y\mid X=0 \sim Bern(\nicefrac{1}{10})$, and $Y\mid X=1 \sim Bern(\nicefrac{8}{10})$. It is clear that the distribution of $Y$ is highly dependent upon the value of $X$. Next note that $Y\sim Bern(p_1)$, where $p_1= \frac{1}{7}\cdot \frac{8}{10} + \frac{6}{7}\cdot \frac{1}{10}= \frac{2}{10}$. Thus, $H(Y)=H(Y\mid X=1)$ and $I_2(X=1;Y)=0$. On the other hand, we have $I_1(X=1;Y)=D(\nicefrac{8}{10}\mid\mid \nicefrac{2}{10})=1.2$ bits. This exemplifies how simply measuring differences in entropy is insufficient for capturing causal influences.
\end{Example}


\subsection{Specific Causal Effects in the Mediation Model}\label{definitions}

Following the process of \cite{pearl2001direct}, we here formalize a series of definitions of total/direct/indirect causal influences from an information theoretic perspective. When leaving the comfort of the unconfounded two-node DAG, it is necessary to incorporate the interventions in the definition of the causal measures:
\begin{Definition}
The specific total effect of $x$ on $Y$ is defined as:
\begin{equation}\label{eq:STE}
STE(x\ra Y)\triangleq \kl{p(Y\mid \hat{x})}{\sum\nolimits_{x'}p(x')p(Y\mid\hat{x}')}
\end{equation}
\end{Definition}
\noindent With the exception of the interventional notation, the STE is equivalent to the SMI. Note that for a DAG given by $X\ra Y$, we will have $STE(x\ra Y)=I_1(x;Y)$ but $STE(y\ra X)=0\ne I_1(y;X)=D(p(X\mid y)\mid\mid p(X))$, where $STE(y\ra X)$ represents the specific total effect of $y$ on $X$. Thus, the STE answers the question posed above in point {\bf(IV)}: ``How much would we expect performing the intervention $do(X=x)$ to change the course of nature for $Y$?''

Next we define the specific controlled direct effect (SCDE) of $x$ on $Y$. Given that computing the controlled direct effect must be done by means of intervention on $Z$, we define the SCDE with respect to a specific value $z$, as it is unclear what distribution over $Z$ should be used if the definition were to take an expectation over \emph{all} possible values of $z$ (see Theorem \ref{prop:exp_cde}).
\begin{Definition}
The specific controlled direct effect of $x$ on $Y$ with mediator $z$ is defined as:
\begin{equation}\label{eq:SCDE}
SCDE(x\ra Y ; z)\triangleq \kl{p(Y\mid \hat{x},\hat{z})}{\sum\nolimits_{x'} p(x')p(Y\mid \hat{x}',\hat{z})}
\end{equation}
\end{Definition}
\noindent The SCDE measures how much we would expect performing the intervention $do(X=x)$ to alter the course of nature for $Y$ given that $Z$ is held fixed at $z$. 

Next, the specific natural direct effect measures the direct effect of $x$ on $Y$ that occurs naturally when the mediator is not controlled:
\begin{Definition}
The specific natural direct effect of $x$ on $Y$ is defined as:
\begin{equation}\label{eq:SNDE}
SNDE(x\ra Y)\triangleq\kl{p(Y\mid \hat{x})}{\sum\nolimits_{x',z'} p(x')p(z'\mid \hat{x})p(Y\mid \hat{x}',z')}
\end{equation}
\end{Definition}

\noindent It is helpful to dissect the two distributions of $Y$ considered by the SNDE. Expanding the first argument as $\sum_{z'}p(z'\mid\hat{x})p(Y\mid \hat{x},z')$, both distributions are given by a weighted combination of the distribution of $Y$ conditioned upon different values of $Z$. In both cases, these values of $Z$ are weighted by the probability with which they would occur under the intervention $\hat{x}$. For the intervened values of $X$ used to evaluate the probability of $Y$, however, the first distribution uses the ``ground truth'' value $x$, whereas the second uses the ``naturally occurring'' $x'$, weighted according to $p(x')$. We can interpret the SNDE as a measure of how much we expect performing the intervention $do(X=x)$ directly alters the course of nature for $Y$. Using the same logic, we can define a specific natural indirect effect:
\begin{Definition}
The specific natural indirect effect of $x$ on $Y$ is defined as:
\begin{equation}\label{eq:SNIE}
SNIE(x\ra Y)\triangleq \kl{p(Y\mid \hat{x})}{\sum\nolimits_{x',z'}p(x')p(z' \mid \hat{x}')p(Y\mid \hat{x},z')}
\end{equation}
\end{Definition}
\noindent Conducting a similar dissection, we see that the roles of $x$ and $x'$ are swapped from the SNDE -- the ``ground truth'' $x$ is used to evaluate the probability of $Y$, while the naturally occurring $x'$ is used to weight different values $z'$. As such, the only difference between the first and second arguments of the SNIE is how the value of the mediating $Z$ is determined, resulting in a measurement of the indirect effect of $x$ on $Y$. We can interpret the SNIE as a measure of how much we expect performing the intervention $do(X=x)$ indirectly alters the course of nature for $Y$.

Unfortunately, the proposed definitions of SNDE and SNIE yield no obvious inequalities with respect to the STE (for example, $SNDE(x\ra Y)+SNIE(x\ra Y)\not\le TE(x\ra Y)$ in general). While this is initially unintuitive, it can be justified by the decision to have all causal influences be assigned a non-negative magnitude. As such, we would expect that contradictory indirect and direct effects could individually have a large magnitude while still resulting in a total effect of zero.

\subsection{Equivalence Relations}\label{properties}
We now analyze the relationship between the proposed specific measures and IF/CS.
\begin{Theorem}\label{prop:exp_te}
The expected STE is equivalent to the information flow, i.e. $\E[STE(X\ra Y)]=I(X\rightarrow Y)$, where the expectation is taken with respect to the marginal distribution over $X$.
\end{Theorem}
\noindent A proof is provided in Appendix \ref{app:exp_te}. The above theorem shows that the expected STE recovers the standard (unconditional) IF from $X$ to $Y$. Notably, the expected STE is \emph{not} equivalent to the CS associated with any subset of the arrows in the graph. Next, we show that both IF and CS provide a notion of expected SCDE:
\begin{Theorem}\label{prop:exp_cde}
The conditional IF is given by the expected value of the SCDE taken with respect to the \emph{marginal} distributions of $X$ and $Z$:
\begin{equation*}
I(X\ra Y\mid \hat{Z})
 = \sum_{x,z} p(x)p(z)SCDE(x\ra Y; z)
\end{equation*}
Furthermore, if the DAG consists of only $X$, $Y$, and $Z$ (i.e. $U=\emptyset$), then the CS of $X\ra Y$ is given by the expected value of the SCDE taken with respect to the \emph{joint} distribution of $X$ and $Z$:
\begin{equation*}
\mathfrak{C}_{X\ra Y}
        = \sum_{x,z} p(x,z)SCDE(x\ra Y; z)
\end{equation*}
\end{Theorem}
\noindent A proof is provided in Appendix \ref{app:exp_cde}. This theorem clarifies the point made earlier with regard to the value of a measure of \emph{natural} direct effect. In particular, when taking an average with respect to possible control values for the mediator $Z$, it is not clear what distribution over $Z$ should be used. 

\subsection{Conditional Specific Influences}\label{sec:setting}
Even though the above causal measures are defined for specific values of $X$, they provide a notion of average causal influence in that they are implicitly averaging over all possible covariates $U$. Given that different values of $u$ may significantly affect the nature of the relationship between $x$ and $Y$, we define conditional versions of the above definitions for a specific value $U=u$. We here consider the general case where only a subset of the covariates $\tilde{U}\subseteq U$ are observed:
\begin{Definition}
The conditional STE of $x$ on $Y$ given $\tilde{u}$ is defined as:
\begin{equation}\label{eq:SSTE}
STE(x\ra Y\mid \tilde{u})\triangleq \kl{p(Y\mid \hat{x},\tilde{u})}{\sum\nolimits_{x'} p(x'\mid \tilde{u})p(Y\mid \hat{x}',\tilde{u})}
\end{equation}
\noindent For the special case where we can observe all relevant covariates, i.e. $\tilde{U}=U$, the conditional STE can be simplified as:
\begin{equation}
STE(x\ra Y\mid u)\triangleq \kl{p(Y\mid \hat{x},u_Y,u_Z)}{\sum\nolimits_{x'}p(x'\mid u_X)p(Y\mid\hat{x}',u_Y,u_Z)}
\end{equation}
\end{Definition}
\noindent This definition violates the locality postulate {\bf (P2)} of Janzing et al. \cite{janzing2013quantifying} in that the causal effect of $x$ on $Y$ may be dependent upon how $X$ is affected by \emph{its own parents}. Allowing this is, however, consistent with the perspective that IT measures quantify the deviance from the course of nature in that the value $u$ dictates the current \emph{natural state}. Nevertheless, the terms $p(x'\mid \tilde{u})$ and $p(x'\mid u_X)$ can be replaced with $p(x')$ if one wishes to remain faithful to the locality postulate (though not explored presently, this would provide us with a notion of \emph{specific} causal strength). The conditional versions of SCDE, SNDE, and SNIE follow very similar logic to that of the STE, and are defined in Appendix \ref{app:settingspecific}.

\subsection{Identifiability}
When $U$ is partially observable or unobservable, the nature of the dependence relationships between $U_X$, $U_Y$, and $U_Z$ will dictate the ability to estimate the proposed causal measures from observational data -- more specifically, the ability to determine the interventional distributions given only estimated conditional distributions. This is crucially important given that performing interventions in many complex natural systems is infeasible. The following theorem uses the d-separation criterion \cite{pearl1988probabilistic,lauritzen1990independence} to identify when the conditional specific measures can be estimated in the partially observable setting where only $\tilde{U}\subset U$ can be observed:
\begin{Theorem}\label{prop:identifiability}
Consider a dataset containing observations of $X$, $Y$, $Z$, and partially observable covariates $\tilde{U}\subseteq U$. Then, the conditional STE, SNDE, and SNIE are non-experimentally identifiable if there exist $\tilde{U}_1,\tilde{U}_2\subseteq\tilde{U}$ such that the following two conditions hold: {\bf(1)} $(X \independent Y \mid \tilde{U}_1)_{\mcal{G}_{\underline{X}}}$ and {\bf(2)} $(X \independent Z \mid \tilde{U}_2)_{\mcal{G}_{\underline{X}}}$, where $\mcal{G}_{\underline{X}}$ represents the DAG with all outgoing arrows from $X$ removed, and $(A \independent B \mid C)_\mcal{G}$ represents the d-separation of $A$ and $B$ by $C$ in DAG $\mcal{G}$.
\end{Theorem}
\noindent The proof uses a direct application of Pearl's $do$-calculus \cite[Theorem 3.4.1]{pearl2009causality}, and is provided in Appendix \ref{app:idendifiability}. By letting $\tilde{U}=\emptyset$, identifiability conditions for the specific unconditional causal effects are obtained. Similarly, the theorem provides the corollary that the conditional specific causal effects may be estimated from observational data when $U$ is fully observable. It is important to note that the above theorem assumes that each conditional distribution can be sufficiently well estimated. Indeed, the ``increased resolution'' of the proposed measures comes at a cost in that reliable estimation of the proposed measures poses challenges for values of $X$ that occur infrequently. Consider, for example, estimating the second argument of the KL-divergence defining the SNDE in \eqref{eq:SNDE}, namely $p(y\mid \hat{x}',z')$. Given that there is a sum over $x'$ and $z'$, it is necessary to know this distribution for \emph{every} pair $(x',z')$. Thus, when $p(x',z')$ is very small, a significant amount of data will be required to estimate $p(y\mid x',z')$ (and therefore the SNDE) reliably.

\subsection{Normalized Specific Effects}
The opacity of measuring causal influences in bits can be addressed by identifying a normalization procedure.
\begin{Definition}
The normalized conditional STE of $x$ on $Y$ conditioned on $\tilde{u}$ is defined as:
\begin{equation}
\overline{STE}(x\ra Y\mid \tilde{u}) \triangleq  \frac{STE(x\ra Y\mid \tilde{u})}{STE(x\ra Y\mid \tilde{u})+H(Y \mid do(X=x),\tilde{U}=\tilde{u})}
\end{equation}
\end{Definition}
\noindent The normalized versions of the other specific causal measures are provided in Appendix \ref{app:normalization}. For the sake of exposition, suppose $\tilde{U}=\emptyset$ and recall the data compression interpretation of $STE(x\ra Y)$ as the excess number of bits used to encode $Y$ under the assumption $X$ occurs naturally when we have in fact forced $X=x$ by means of an intervention. Noting that $H(Y \mid do(X=x))$ represents the number of bits required to encode $Y$ when we have (knowingly) forced $X=x$, the denominator of $\overline{STE}$ gives the \emph{total} number of bits used to encode $Y$ under the incorrect assumption of a naturally occurring $X$. As such, the normalized STE represents the fraction of bits used to encode $Y$ under the assumption that $X$ occurred naturally that are unnecessary when performing the intervention $do(X=x)$. 

As a result of the non-negativity of entropy and the KL-divergence, the normalized STE is bounded between zero and one. Interpreting $\overline{STE}$ is facilitated by considering the scenarios that yield the extremal values. First, the normalized STE is zero if and only if the STE is zero, which is to say that $p(y\mid \hat{x},\tilde{u})=p(y\mid \tilde{u})$ for all $y$ for which $p(y\mid \hat{x},\tilde{u})>0$. More interestingly, the normalized STE is one if and only if the STE is greater than zero and $H(Y\mid do(X=x),\tilde{U}=\tilde{u})=0$. As such, the normalized STE being equal to one represents $x$ having a maximal causal effect on $Y$ in the sense that performing the intervention $do(X=x)$ determines the value of $Y$ with 100 percent certainty. It should be emphasized that, like the unnormalized measures, this notion of maximal causal effect applies strictly in a distributional sense and says nothing of the direction or magnitude of the causal effect with respect to the units of $Y$. For example, if performing $do(X=x)$ results in $Y=\E[Y]$ with probability one, then $H(Y\mid do(X=x))=0$ and we would conclude that $x$ has a maximal effect on $Y$ even though $x$ causes $Y$ to take the value it is \emph{expected} to take absent an intervention.

\section{Examples}\label{examples}

We now present three examples of notions of causal influence that are uniquely identified by the specific causal measures.

\subsection{Chain Reaction}\label{sec:chain_reaction}
For the first example consider a simple chain $X\ra Z \ra Y$. This can be thought of as a simplified version of the example proposed by Ay and Polani \cite{ay2008information} and modified to include noise by Janzing et al. \cite[Example 7]{janzing2013quantifying}. We will consider the simplest case of this example where a binary message is being passed from $X$ to $Z$ to $Y$, with the message being flipped by $Z$ and $Y$ with probability $\epsilon$. We will interpret each variable as representing the message it passes on, i.e. $X=1$ means ``$X$ passes the message $1$ to $Z$.'' Formally, let $X,Y,Z\in\{0,1\}$ with $X\sim Bern(0.5)$:
\begin{equation}
Z=\begin{cases}
X &w.p. \ 1-\epsilon \\
X \oplus 1 &w.p. \ \epsilon
\end{cases} \ \ \ \ \ \ \ \ \ \
Y=\begin{cases}
Z &w.p. \ 1-\epsilon \\
Z \oplus 1 &w.p. \ \epsilon
\end{cases}
\end{equation}
where $\oplus$ is the XOR operation.

Focusing first on the effect of $x$ on $Y$, we note that because the only path from $X$ to $Y$ is the one through $Z$, the direct effect is zero and the total and indirect effects are equal. Noting that $Y\sim Bern(0.5)$, $Y\mid do(X=0)\sim Bern(2\epsilon(1-\epsilon))$, and $Y\mid do(X=1)\sim Bern(1-2\epsilon(1-\epsilon))$, the total effect is the same for both $x\in \{0,1\}$ and is given by:
\begin{align}
STE(x\ra Y) = D(2\epsilon(1-\epsilon)\mid\mid 0.5) \xrightarrow[\epsilon\ra 0]{} 1
\end{align}
\noindent Thus, as the probability of flipping the message approaches zero, $Y$ will be deterministically linked to $X$, and $X$ resolves the entire one bit of uncertainty associated with $Y$. Now consider the conditional STE of $z$ on $Y$ for a particular $x$. We can compute this by comparing the distributions $p(y\mid x,\hat{z})=p(y\mid z)$ and $p(y \mid x)$. Given the symmetry of the problem, this will take one of two values depending on whether or not $x$ and $z$ are equal:
\begin{equation*}
STE(z\ra Y\mid x) =
\begin{cases}
\kl{\epsilon}{2\epsilon(1-\epsilon)} &x=z\\
\kl{\epsilon}{\epsilon^2 + (1-\epsilon)^2} &x\ne z
\end{cases}
\end{equation*}
As $\epsilon$ approaches zero, the STE approaches zero when $x=z$ and infinity when $x\ne z$. To understand this result, fix $\epsilon$ to be an arbitrarily small number such that $Z$ will pass on its received message with high probability. Thus, when $x=z$, it is, in a sense, unreasonable to endow $Z$ with responsibility for causing the value taken by $Y$ when it is propagating the message in a nearly deterministic manner. In such a case, it is not so much $Z$ that is causing $Y$, but rather $X$ that initiated a \emph{chain reaction}. On the other hand, in the unlikely occurrence that $x\ne z$, we have that $Z$ does have a causal effect on $Y$. This scenario can be thought of as $Z$ acting of its own volition in selecting a message to pass to $Y$.

We acknowledge that the notion of an unbounded causal influence is initially unsettling. When looking closer, however, this property is intuitive. First, we note that for any fixed $\epsilon >0$, the STE will be finite. It is only for $\epsilon=0$ that the STE could be infinite, but in that case, the setting that results in infinite influence happens \emph{with probability zero}. Thus, in general, an infinite influence could only be achieved through intervention. Furthermore, such an intervention would have to assign a value to a cause that occurs with probability zero, and that cause would in turn have to enable an otherwise impossible effect to have non-zero probability.

This conditional formulation violates the locality postulate {\bf (P2)} of Janzing et al. \cite{janzing2013quantifying} in that the effect of $z$ depends on the value of its own parent, $x$. We do not claim that the perspective taken here is ``correct,'' but merely point out that there exist justifications for considering the value of a cause's parent in evaluating the causal effect.

\subsection{Caused Uncertainty}\label{uncertain}
Consider a 3-node DAG characterized by the connections $X\ra Y\leftarrow Z$ with $X\sim Bern(0.5)$, $Z\sim Bern(0.1)$ and:
\begin{equation*}
Y\mid X,Z\sim \begin{cases}
Bern(0.5) & Z=1 \\
Bern(0.1) & (X,Z)=(0,0) \\
Bern(0.9) & (X,Z)=(1,0)
\end{cases}
\end{equation*}

\noindent Given that $X$ and $Z$ are both parentless, we can treat interventions on $X$ and $Z$ as observations, and the CS, conditional IF, and conditional mutual information (CMI) are equivalent. In particular, we have that $\mathfrak{C}_{X\ra Y}=I(X\ra Y\mid \hat{Z})=I(X;Y\mid Z)\approx 0.48$ and $\mathfrak{C}_{Z\ra Y}=I(Z\ra Y\mid \hat{X})=I(Z;Y\mid X)\approx 0.06$. Writing CMI as a difference of conditional entropies $I(Z;Y\mid X)=H(Y\mid X) - H(Y\mid X,Z)$ provides us with the interpretation of CMI as the reduction in uncertainty of $Y$ resulting from the added conditioning of $Z$, which will always be non-negative.

Next we consider $STE(x\ra Y\mid z)$ and $STE(z\ra Y\mid x)$ for $(x,z)\in \{0,1\}^2$. Given the symmetry of the problem with respect to $X$, we only need to consider two of the four possible values of $(X,Z)$, namely $(x_0,z_0)\triangleq(0,0)$ and $(x_0,z_1)\triangleq(0,1)$. In order to compute the STE for each $X$ and $Z$ to $Y$ in either case, we need the following distributions:
\begin{align*}
&p(Y\mid x_0,z_0) = Bern(0.1)
&&p(Y\mid x_0,z_1) = Bern(0.5) \\
&p(Y\mid z_0) = Bern(0.5)
&&p(Y\mid z_1) = Bern(0.5) \\
&p(Y\mid x_0) = Bern(0.14)
&&
\end{align*}
For a given $(x,z)$, the STE is given by $STE(x\ra Y\mid z)=\kl{p(Y\mid x,z)}{p(Y\mid z)}$ and $STE(z\ra Y\mid x)=\kl{p(Y\mid x,z)}{p(Y\mid x)}$:
\begin{equation*}
STE(x\ra Y\mid z)\approx
\begin{cases}
0.53 &z=0\\
0.00 &z=1
\end{cases}
\ \ \ \ \ \ \ \
STE(z\ra Y\mid x)\approx
\begin{cases}
0.01 &z=0\\
0.52 &z=1
\end{cases}
\end{equation*}
The results presented above are intuitive: when $z=0$, then the value taken by $Y$ is largely determined by $X$, and the knowledge that $z=0$ tells us very little about the distribution of $Y$. On the other hand, when $z=1$, $X$ has no bearing on the value taken by $Y$. Thus, in this scenario, it is the value taken by $Z$ that has caused the shift in the distribution of $Y$, even though $Z$ provides no information with regard to the particular \emph{value} taken by $Y$. In this sense, we can think of $Z$ as \emph{causing uncertainty} in $Y$. This scenario makes particularly clear why it makes sense to condition on the cause but take an expectation with respect to the effect -- no outcome $y$ could be attributed to being a result of $z=1$, despite the clear influence that such an event has on the distribution of $Y$.

\subsection{Shared Responsibility}
Consider a scenario where a collection of $n$ iid variables $X_i\sim Bern(\epsilon)$ collectively influence a single outcome $Y$, i.e. $X_i\ra Y$ for $i=1,\dots,n$. For a given context $\{x_i\}_{i=1}^n$, let $k$ be the number of $x_i$ that are one, i.e. $k=\sum_i x_i$. Then let $Y$ be distributed as:
\begin{equation*}
Y\mid X_1,\dots,X_n\sim
Bern\left(\frac{1}{2^K}\right)
\end{equation*}
where $K=\sum_i X_i$ is a random variable. One interpretation of this example is that each $X_i$ is a potential inhibitor of $Y$. As more inhibitors become activated (i.e. as $k$ grows), the effect of adding another inhibitor diminishes. Since the value taken by $K$ depends on the values taken by each $X_i$, a measure that averages with respect to $X_i$ will not capture this change in causal effect that results for different values of $k$.

As with the previous example, the CS, conditional IF, and CMI are equivalent for this problem setting. While there is no simple computation for these measures as a function of $\epsilon$ and $n$, there are a couple of key points. First, the influence of each of the variables $X_i$ on $Y$ is the same, i.e. $I(X_i;Y\mid X_1,\dots,X_{i-1},X_{i+1},\dots,X_n) = I(X_1;Y\mid X_2,\dots,X_n)$ for all $i=1,\dots,n$. Second, as $n\ra \infty$, the probability of $Y=1$ goes to zero, and as $\epsilon\ra 0$, the probability of $Y=1$ goes to one. In either of the limits, the entropy of $Y$ goes to zero and thus so does the causal influence of each $X_i$ as measured by either CMI, conditional IF, or CS.

Now consider a realization $\{x_i\}_{i=1}^n$ and the corresponding $STE(x_1\ra Y\mid x_2,\dots,x_n)$. While the influence of each $x_i$ on $Y$ will \emph{not} be the same for a given realization, the symmetry of the problem is such that the computation will be performed in the same manner for each $x_i$. Letting $k_1\triangleq\sum_{i=2}^n x_i$ be the number of ones excluding $x_1$, define the following distributions:
\begin{gather*}
p(Y\mid \{x_i\}_{i=1}^n)=
p(Y\mid k)=
Bern\left(\frac{1}{2^k}\right) \\
p(Y\mid \{x_i\}_{i=2}^n)=
p(Y\mid k_1)=
Bern\left(\frac{\epsilon}{2^{k_1+1}}+\frac{1-\epsilon}{2^{k_1}}\right)
\end{gather*}
Then, for a given realization, the STE is a function of $x_1$ and $k_1$:
\begin{align*}
STE(x_1\ra Y\mid k_1)&=
\kl{p(Y\mid k)}{p(Y\mid k_1)}\\
&=\begin{cases}
\kl{\frac{1}{2^{k_1}}}{\frac{\epsilon}{2^{k_1+1}}+\frac{1-\epsilon}{2^{k_1}}} &x_1=0 \\
\kl{\frac{1}{2^{k_1+1}}}{\frac{\epsilon}{2^{k_1+1}}+\frac{1-\epsilon}{2^{k_1}}} &x_1=1
\end{cases}
\end{align*}
In interpreting these results, first assume that $\epsilon$ is small, meaning that for each of the inhibitors, it is unlikely that it will be activated. As a result of this assumption, we have $STE(X_1=0\ra Y\mid k_1) < STE(X_1=1\ra Y\mid k_1)$, i.e. an inhibitor has a greater influence when it is activated. More interestingly, note that $STE(x_1\ra Y\mid k_1)$ is strictly decreasing in $k_1$. This is consistent with the intuition provided above, namely that if a large number of inhibitors are active, then they \emph{share responsibility} and the influence of any single one is negligible. On the other hand, if only one is activated (i.e. $(x_1,k_1)=(1,0)$), then in the limit of $\epsilon\ra 0$, its influence will approach infinity (and its normalized influence will approach one).


\section{Case Study -- Effect of El Ni\~no-Southern Oscillation on Pacific Northwest Temperature Anomalies}\label{sec:climate}

\begin{figure*}[t!]
  \begin{center}
    \includegraphics[width=0.9\linewidth]{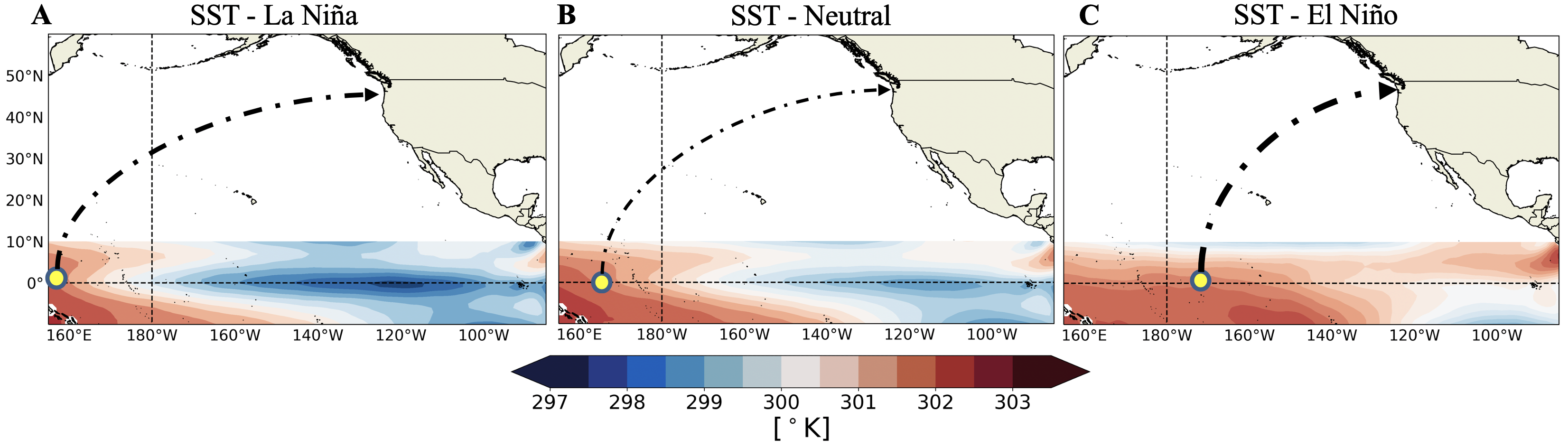}
  \end{center}
  \caption{Sea surface temperatures (SST) averaged over January, February, and March from 1979-2018 in the equatorial Pacific for La Ni\~na (A), neutral (B), and El Ni\~no (C) ENSO phases derived from the ERA-interim OCEAN5 reanalysis product conditioned on the Ni\~no 3.4 index $\pm$ 1 anomaly standard deviation \cite{zuo2019ecmwf}. The shifted SST patterns give rise to shifted precipitation regions (yellow circle), which affect temperature anomalies in the PNW through large scale atmospheric waves.}\label{fig:nnn}
\end{figure*}
We now present an application of the proposed framework to measuring the specific causal influences of the El Ni\~no-Southern Oscillation (ENSO) on the temperature anomaly signal in the North American Pacific Northwest (PNW, latitude: 47$^\circ$N, longitude: 240$^\circ$E). The dataset we use is publicly available at the National Center for Atmospheric Research website and all code is published on the Code Ocean platform\footnote{\url{https://doi.org/10.24433/CO.5484914.v1}}. For our purposes, ENSO is characterized by the sea surface temperature in the Ni\~no 3.4 region located in the equatorial Pacific (latitude: 5$^\circ$S-5$^\circ$N, longitude: 120$^\circ$W-170$^\circ$W). The ENSO signal is typically understood by being in one of three phases (or states) -- a neutral phase (we will refer to this as $E=0$) gives rise to a precipitation region centered near longitude 160$^\circ$E (Figure \ref{fig:nnn}B), the El Ni\~no phase ($E=1$) gives rise to an eastward shifted precipitation region ($\sim$170$^\circ$W, Figure \ref{fig:nnn}C), and the La Ni\~na phase ($E=-1$) gives rise to a westward shifted precipitation region ($\sim$150$^\circ$E, Figure \ref{fig:nnn}A) \cite{alexander2002atmospheric,henderson2018enso}. Ni\~no and Ni\~na phases can occur with varying intensities during the winter months with a typical return period of two to seven years \cite{li2013nino}. When a Ni\~no or Ni\~na phase occurs, the shifted precipitation signal produces large scale atmospheric Rossby waves (waves in the upper level atmospheric pressure field) that influence North American land temperatures, predominantly through the well studied Pacific North American teleconnection pattern (PNA) \cite{bjerknes1969atmospheric,wallace1981teleconnections}. PNA affects North American land temperatures through the advection of warm marine air during a Ni\~no phase and cool polar air during a Ni\~na phase \cite{zhou2014global,hoerling1997nino}. We here use the proposed framework to quantify the causal effect of this teleconnection, focusing specifically on the temperature in the PNW.

This application is a particularly good fit for the proposed analysis for a number of reasons. First, by utilizing a collection of simulation model runs, an immense amount of data can be obtained. Second, domain expertise can be leveraged to construct causal DAGs prior to performing analysis. For example, it is well known that the ENSO signal influences temperature as opposed to the temperature influencing ENSO. Third, there are well-accepted methods for detrending signals, and these methods can be used to control for possible confounding effects. Fourth, it is to be expected that different phases of the ENSO signal will, in some sense, give rise to larger causal effects than other phases \cite{an2004nonlinearity}. The proposed framework can be used to quantify these differences in a formal sense.


\begin{figure}[t!]
  \begin{center}
    \includegraphics[width=\textwidth]{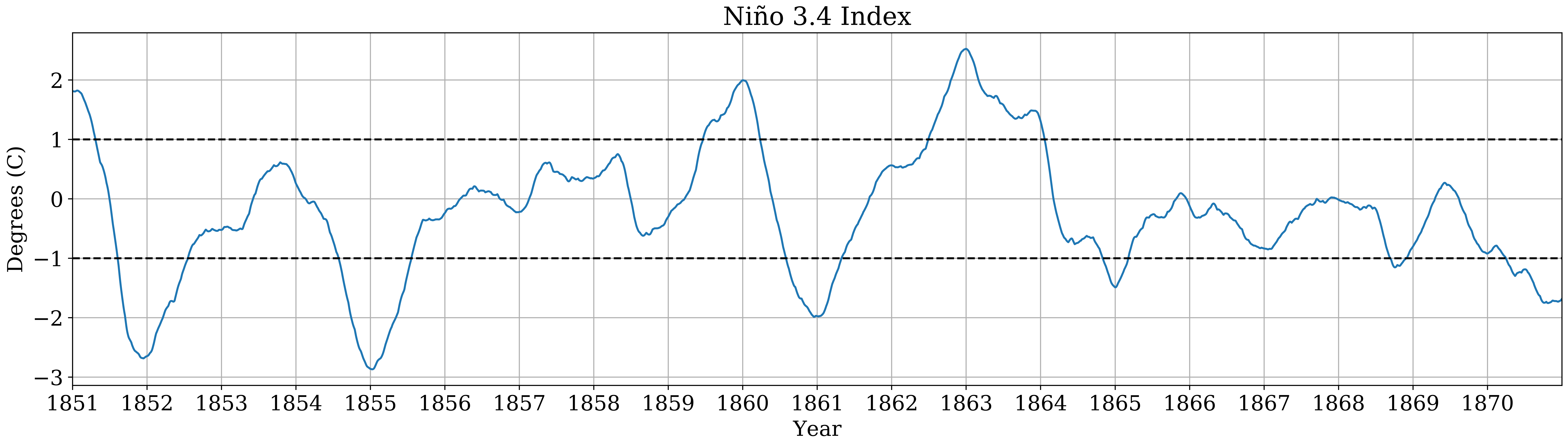}
  \end{center}
  \caption[Ni\~no 3.4 index from 1851-1871.]{Simulation of the Ni\~no 3.4 index from 1851-1871 from a CESM2 model run along with threshold for determining ENSO phase.}\label{fig:enso}
\end{figure}

The analyzed dataset is composed of nine simulated model runs from the National Center for Atmospheric Research's (NCAR) Community Earth System Model, version 2 (CESM2) \cite{gettelman2018regional} scientifically validated historical CMIP6 runs \cite{eyring2016overview}. Each model run provides an array of daily temperature values spanning the years 1850 to 2015 from which we can compute the Ni\~no 3.4 index (as in \cite{bamston1997documentation}) and directly obtain the PNW two-meter temperature. The Ni\~no 3.4 index is a measure of anomalous equatorial sea surface temperatures in the Ni\~no 3.4 region described above. Each of the model runs provides an independent realization of possible evolutions of temperatures that obey the underlying dynamic and thermodynamic equations as encoded by the model. It is important to clarify that the model is not intended for prediction, but rather gives possible atmospheric states for a given set of initial conditions and constraints determined by the selected time period (i.e. CO$_2$ forcing, solar/lunar cycles, etc.). Both the ENSO index and PNW two-meter temperature signals have the mean and the leading six harmonics of the annual cycle removed, leaving only the anomalous components of the signal. As this is standard practice in the analysis of climate data (e.g. \cite{krahmann2001formation}), we henceforth strictly consider anomaly signals.

A 20-year model run of the Ni\~no 3.4 index is shown in Figure \ref{fig:enso}. It is clear that the ENSO signal does not reliably alternate between $E=1$ and $E=-1$ with a constant period. As a result of ENSO cold-season phase locking \cite{neelin2000variations}, the ENSO signal is strongest in or near to January (marked by vertical grid lines). As such, we limit our focus to the months of January, February, and March, as it is not interesting to measure the effect of the ENSO signal in the months where it is not present. We further simplify the problem by quantizing the ENSO index on an annual timescale, i.e. we assign a single value to $E\in\{-1,0,1\}$ for January-March of a given year based on the ENSO index value on January 1st of that year.

Given that we are estimating the effect of ENSO on temperature, we similarly consider the temperature signal only during the months of January, February, and March. Rather than attempting to assess the effect of ENSO on daily temperature anomalies, we choose to focus on two-week averages, corresponding to the limit of predictability in numerical weather forecasting \cite{magnusson2013factors}. As we will discuss in the next section, this choice also facilitates the causal modeling. As a final processing step, we quantize the temperature anomaly averages to $T \in \{-1,0,1\}$. While this quantization does come with an inevitable loss of resolution, it yields the easily understood interpretation of the temperature signal as representing either a cold anomaly, a warm anomaly, or neutral state. We compute the quantization threshold on the entire dataset (i.e. before averaging and before selecting for months) such that one third of days are in each category. The averages are then compared to these thresholds, given by -1.3 and +1.94 degrees Kelvin. The resultant dataset after selecting for the winter months and taking two-week averages consists of 9840 samples.
\subsection{Causal Model}
\begin{figure}[t!]
  \begin{center}
    \includegraphics[width=0.7\linewidth]{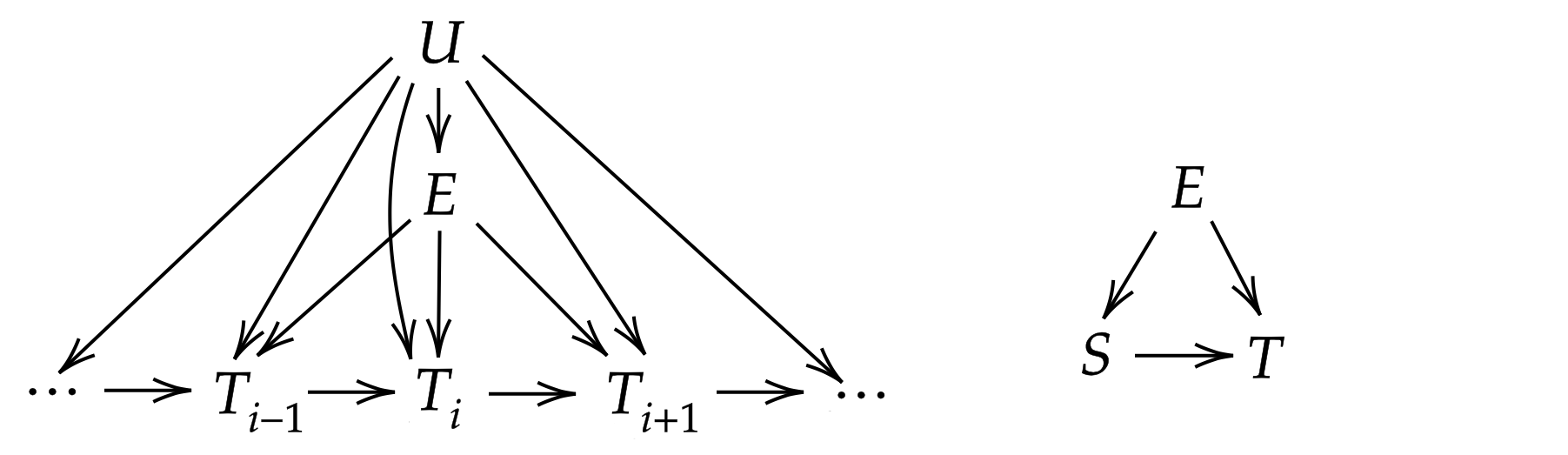}
  \end{center}
  \caption[Causal DAG representations of climate variables.]{Left: Complete DAG representation of climate variables. Right: Simplified DAG after detrending and assuming stationarity.}\label{fig:climate_dags}
\end{figure}

In order to implement the proposed framework, we first need to formulate a causal DAG representation of the dataset discussed above. As a starting point, consider the DAG on the left side of Figure \ref{fig:climate_dags}, where we let $E$ represent an annual ENSO phase, $T_1,\dots,T_6$ represent the quantized two-week temperature anomaly averages for January through March (i.e. $T_1$ averages January 1st through 14th, $T_2$ averages January 15th through 28th, etc.), and $U$ represents the other factors, such as seasonality and CO$_2$ forcing. This DAG encodes a number of assumptions. First, it encodes the intuition that seasonality may affect ENSO and the temperature, but not the other way around. Similarly, ENSO will affect the temperature in the PNW, but not the other way around. The more interesting implicit assumption is that there is a persistence signal in the temperature represented by the arrow $T_{i-1}\ra T_i$. Importantly, we have assumed that this persistence signal is Markov (when conditioned on $E$ and $U$), i.e. there is no arrow $T_{i-k}\ra T_i$ for $k>1$. This assumption significantly simplifies estimation of the direct and indirect effects of $E$ on $T_i$, as those require estimating the distribution of $T_i$ for every possible combination of its parents. This serves as a motivation for the decision to consider two-week averages -- if we were to simply consider daily temperatures, it is unreasonable to expect that $T_i$ would be independent of $T_{i-2}$ when conditioned on $E$, $U$, and $T_{i-1}$.

We next incorporate two assumptions in order to simplify the causal model. First, we assume that all the effects of $U$ are removed by the detrending and removal of annual cycle performed in the preprocessing steps. It is to be expected that this assumption will hold for the well known shared causes (such as the aforementioned seasonality and CO$_2$ forcing), but the possibility of other factors that have effects not captured by the leading six harmonics of the annual cycle is important to note. The second assumption we make is that the distribution of the temperature anomaly averages does not change over time, i.e. that $p(t_i \mid t_{i-1},e)$ and $p(t_i \mid e)$ are not dependent on $i$. After making these assumptions, we obtain the simplified DAG on the right of Figure \ref{fig:climate_dags}, where we introduce the new variable $S$ to represent the past temperature anomaly average and $T$ to represent the subsequent temperature average, and note that this perfectly matches the mediation model in Figure \ref{fig:dag} with $U=\emptyset$. We can think of $T$ as representing $T_i$ and $S$ as representing either $T_{i-1}$ or the collection $T_1,\dots, T_{i-1}$. To see that these interpretations of $S$ are equivalent, consider the SNDE, given by:
\begin{equation}\label{snde_ex}
SNDE(e\ra T)=D(
p(T\mid \hat{e})\mid\mid
\sum\nolimits_{e',s'}p(e')p(s'\mid \hat{e}')p(T\mid \hat{e},s'))
\end{equation}
Now let $T=T_i$ and $S=T_1,\dots,T_{i-1}\triangleq T_1^{i-1}$, and note that:
\begin{gather*}
p(s\mid \hat{e})=p(t_1^{i-1} \mid \hat{e})=p(t_{i-1}\mid \hat{e})p(t_1^{i-2}\mid \hat{e},t_{i-1}) \\
p(T\mid \hat{e},s)=p(T\mid\hat{e},t_1^{i-1})=p(T\mid\hat{e},t_{i-1})
\end{gather*}
Plugging these into the second argument of the KL-divergence in \eqref{snde_ex}, we get:
\begin{align*}
\sum_{e',s'}p(e')p(s'\mid \hat{e}')p(T\mid \hat{e},s')&=\sum_{e',{t_1^{i-1}}'}p(e')p(t_{i-1}'\mid \hat{e}')p({t_1^{i-2}}'\mid \hat{e}',t_{i-1}')p(T\mid\hat{e},t_{i-1}') \\
&=\sum_{e',t_{i-1}'}p(e')p(t_{i-1}'\mid \hat{e}')p(T\mid\hat{e},t_{i-1}') \left[\sum_{{t_1^{i-2}}'} p({t_1^{i-2}}'\mid \hat{e}',t_{i-1}')\right] \\
&=\sum_{e',t_{i-1}'}p(e')p(t_{i-1}'\mid \hat{e}')p(T\mid\hat{e},t_{i-1}')
\end{align*}
Given that $S$ appears nowhere in the first argument the KL-divergence, we can see that whether $S=T_{i-1}$ or $S=T_1^{i-1}$, the result is the same. The same procedure can be applied to show equivalence for the SNIE. We here choose the interpretation $S=T_{i-1}$. As a result of the assumption that $p(t_i\mid e)$ does not depend on $i$, we have that $p(t\mid e)= p(s\mid e)$ for $t=s$. It should be noted that for $T=T_1$ (i.e. the average for the first two weeks of January), we define $S=T_0$ to be the average taken over the last two weeks of December.

\subsection{Estimation and Significance Testing}

\begin{figure}[t!]
  \begin{center}
    \includegraphics[width=0.85\textwidth]{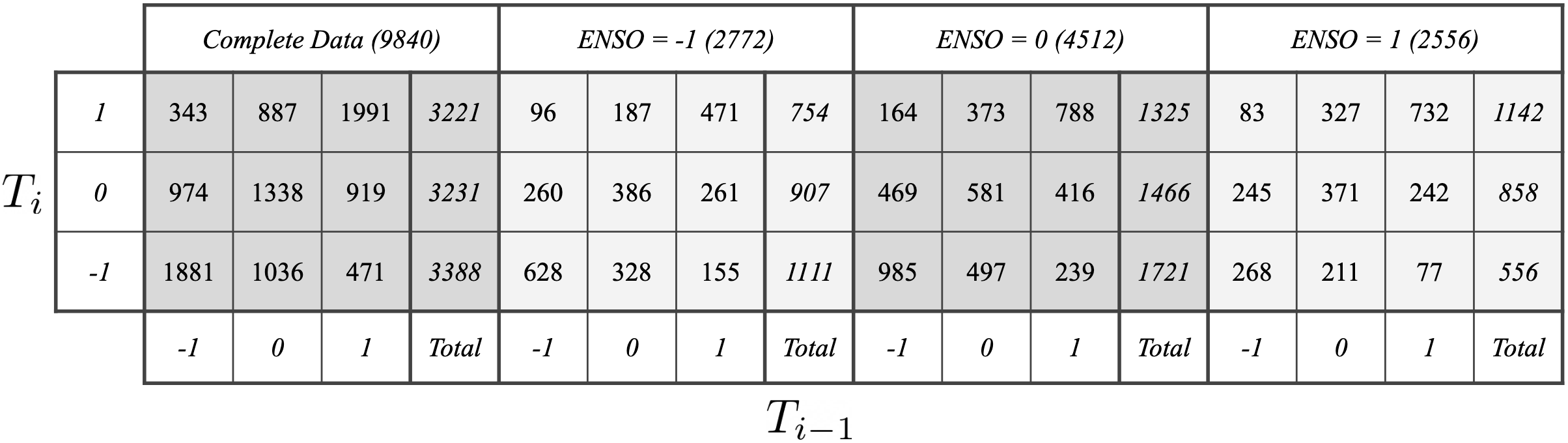}
  \end{center}
  \caption[Number of temperature average transitions]{Counts of transitions from the past average temperature $T_{i-1}$ to the current average $T_i$ in the complete dataset and subsets corresponding with specific values of the ENSO signal. The parenthetical gives the total count of samples in a given subset. The number of samples used to estimate the distribution of the current temperature anomaly for a given ENSO phase and past temperature are given by the sums of the columns. It is clear that there is ample data for estimating each distribution in question (see \eqref{num_samples}).}\label{fig:temp_data}
\end{figure}

We define the dataset from which we estimate the causal influences as $\mc{D}=(e_n,s_n,t_n)_{n=1}^{9840}$. Given that there is a large amount of data and a relatively small alphabet size, we utilize plug-in estimators of the proposed measures, where every distribution in question is estimated using a maximum likelihood estimator. Since $E$ has no parents reduced DAG, we can freely exchange interventions $\hat{e}$ for observations $e$ in the estimation of the effect of $e$ on $T$. As such, the estimates of the specific effect of ENSO on temperature are given by:
\begin{align*}
\widehat{STE}_\mc{D}(e\ra T)&\triangleq\kl
{\hat{p}_\mc{D}(T\mid e)}
{\hat{p}_\mc{D}(T)} \\
\widehat{SNDE}_\mc{D}(e\ra T)&\triangleq\kl
{\hat{p}_\mc{D}(T\mid e)}
{\sum\nolimits_{e',s'} \hat{p}_\mc{D}(e')\hat{p}_\mc{D}(s'\mid e)\hat{p}_\mc{D}(T\mid e',s')} \\
\widehat{SNIE}_\mc{D}(e\ra T)&\triangleq\kl
{\hat{p}_\mc{D}(T\mid e)}
{\sum\nolimits_{e',s'} \hat{p}_\mc{D}(e')\hat{p}_\mc{D}(s'\mid e')\hat{p}_\mc{D}(T\mid e,s')}
\end{align*}
\noindent where $\hat{p}_\mc{D}$ gives the maximum likelihood estimate of $p$ on the sample $\mc{D}$ (see Appendix \ref{app:estimation}). 

Next note that the conditional STE of the past temperature average $S$ on the subsequent temperature $T$ conditioned on an ENSO state $E$ is:
\begin{equation}
STE(s\ra T\mid e) = \kl
{p(T\mid \hat{s},e)}
{\sum\nolimits_{s'}p(s'\mid e)p(T\mid \hat{s}',e)}
\end{equation}
Letting $X=S$, $Y=T$, $Z=\emptyset$, and $U=E$, it follows from Theorem 3 that we can estimate the total effect from observational data. Therefore, we use the following plug-in estimator:
\begin{equation}
\widehat{STE}_\mc{D}(s\ra T\mid e)\triangleq\kl
{\hat{p}_\mc{D}(T\mid e,s)}
{\hat{p}_\mc{D}(T\mid e)}
\end{equation}
Given the absence of an intuitive link between bits and temperature, we choose to focus on the normalized versions of the proposed causal measures (see Appendix \ref{app:estimation}).

By applying these estimators to the complete dataset \mc{D}, we obtain point estimates of the desired measures. For ease of notation, we omit $\mc{D}$ from the estimates from here on. It is important to note that even though not all estimates will utilize all $9840$ samples, Figure \ref{fig:temp_data} makes clear there is a considerable amount of samples available for estimating every distribution in question. In particular, we see that:
\begin{align}
\min_{e,s}&\left|\{n:e_n=e,s_n=s\}\right|=\left|\{n:e_n=1,s_n=-1\}\right|\label{num_samples}\\
&=\sum_t\left|\{n:t_n=t,e_n=1,s_n=-1\}\right|=83+245+268=596 \nonumber
\end{align}
In other words, the distribution estimated on the smallest number of samples is $p(t\mid E=1,S=-1)$, and this estimate is obtained from 596 samples.

In addition to these point estimates, it is desirable to have a means of measuring the significance of the estimated measures and quantifying the uncertainty in our estimates. To achieve these goals, we perform a nonparametric bootstrap hypothesis test \cite{mackinnon2009bootstrap} and construct a nonparametric bootstrap confidence interval \cite{efron1987better}. The goal of the hypothesis test is to estimate the distribution of the estimated measure under a null hypothesis (H0) and assess the likelihood that our estimate came from such a distribution. In this case, H0 corresponds to the absence of a causal link, which would result in the true causal measure being equal to zero. The primary challenge to performing this test is the generation of samples from a distribution representative of H0. We accomplish this using a scheme similar to that presented in \cite[Example 2]{janzing2013quantifying} wherein we group the data by one of the three variables ($E$, $S$, or $T$) and shuffle the other two in order to break one of the causal links. For example, when performing the test for the direct effect of $E$ on $T$, we split the data into three sets: $\{n_{-1}:s_n=-1\}$,  $\{n_0:s_n=0\}$, and  $\{n_{1}:s_n=1\}$. Within each of these sets, we shuffle (i.e. permute) all the samples of $E$ (or $T$). Because the shuffling occurs within groupings of $S$, any possible link from $E$ to $S$ and $S$ to $T$ is preserved (and thus so is the indirect effect), but the link between $E$ and $T$ is destroyed. Each of these permutations is then treated as a sample under H0 from which we estimate the SNDE. We perform this shuffling and estimation procedure 10000 times and use the 95th percentile as the cutoff threshold for statistical significance. This threshold is given by the upper whisker on the boxplots labeled H0 in the figures in the next section. When performing this test for the indirect effect, we choose to break the link from $S$ to $T$ rather than from $E$ to $S$ in order to preserve the assumption that $p(s\mid e)=p(t \mid e)$ for $s=t$.

To quantify the uncertainty in our point estimate we construct a nonparametric bootstrap confidence interval by repeatedly drawing a collection of samples from the empirical distribution of our data and estimating the measure on the new collection of samples. Specifically, let $\mc{D}^*_b=(e_{j^n_b},s_{j^n_b},t_{j^n_b})_{n=1}^{9840}$ be the $b$th bootstrap sample, where $j^n_b$ are drawn independently from the uniform distribution over $\{1,2,\dots,9840\}$ for $b=1,\dots,10000$. We estimate the causal measure in question on each of the 10000 bootstrap samples and use the 5th and 95th percentiles as the lower and upper bounds of the confidence interval.

\subsection{Results}
\begin{figure}
  \begin{center}
    \includegraphics[width=\linewidth]{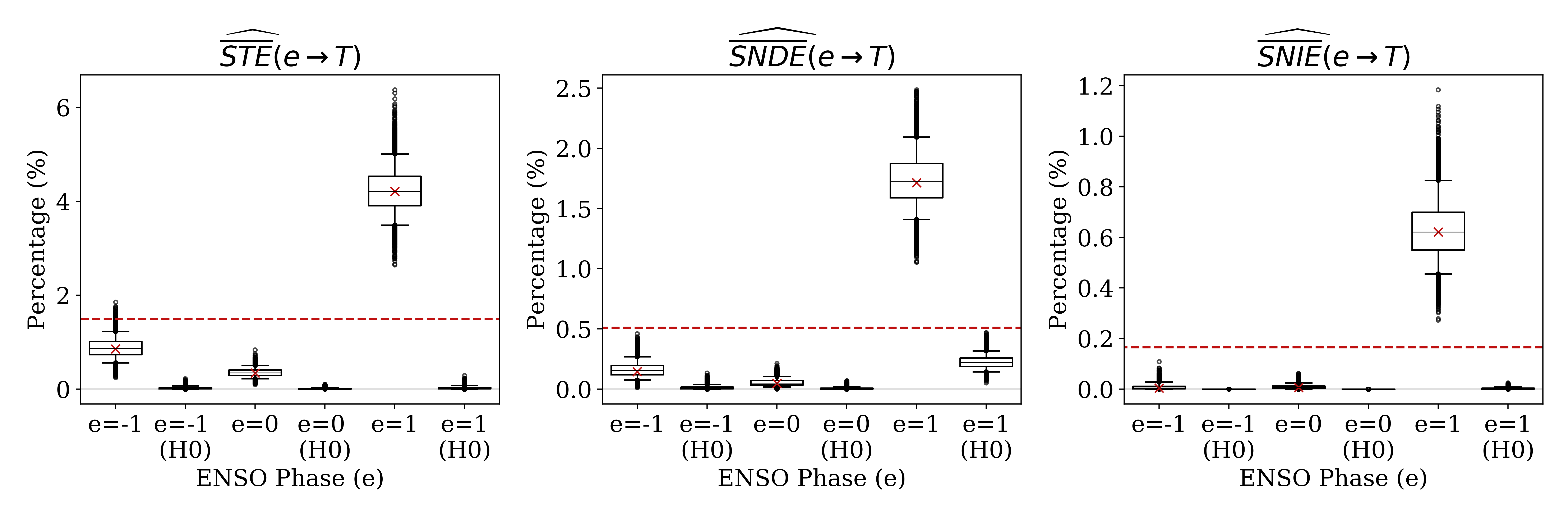}
  \end{center}
  \caption[Specific causal effects of ENSO on temperature anomaly.]{ Estimates of the normalized specific total effect (left), specific natural direct effect (center), and specific natural indirect effect (right) of ENSO on temperature anomalies.}\label{fig:E2T}
\end{figure}

\begin{figure}
  \begin{center}
    \includegraphics[width=\linewidth]{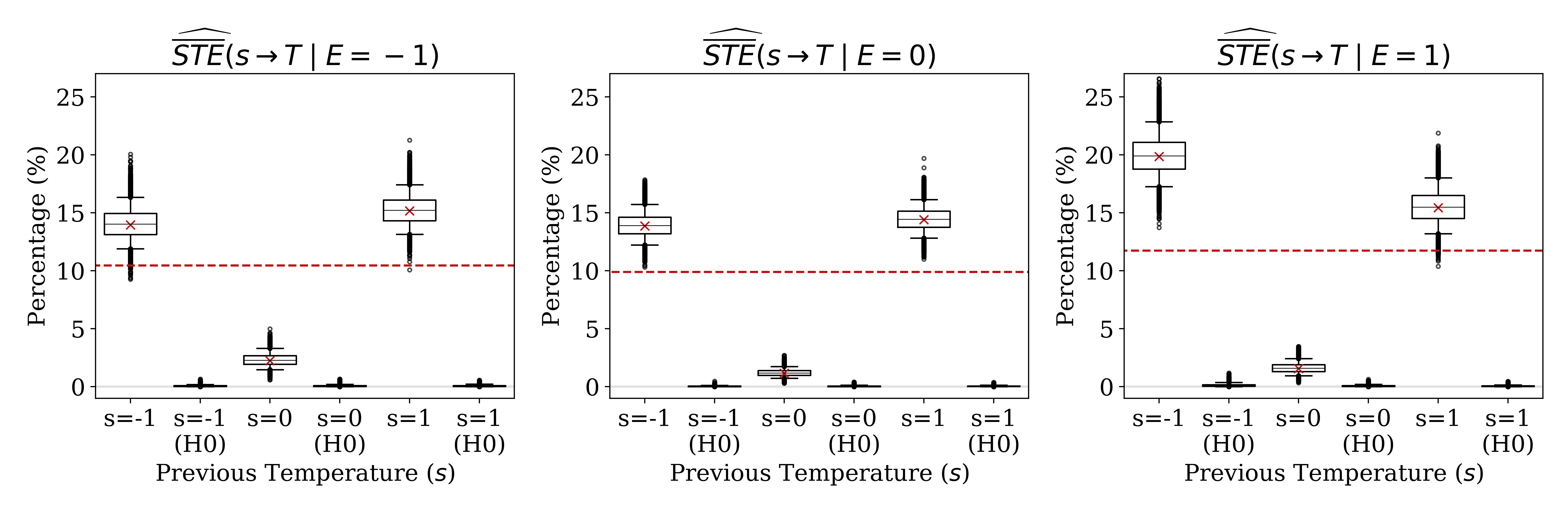}
  \end{center}
  \caption{Estimates of the normalized conditional specific total effect of previous temperature anomaly on current temperature anomaly conditioned on different values of ENSO phase.}\label{fig:S2T}
\end{figure}

We estimate the normalized STE, SNDE, and SNIE of ENSO on temperature and the normalized conditional STE of the past temperature average on the next average conditioned on ENSO. In every case, the measure is estimated on the complete dataset (red $\times$) and compared with the corresponding weighted average (i.e. ``non-specific'') measure (red dashed lines). For the specific measure, we obtain an estimate for each value of the cause, i.e. $e\in\{-1,0,1\}$ or $s\in \{-1,0,1\}$. The average measure is then calculated by taking an expectation of the specific measures with respect to $\hat{p}(e)$ or $\hat{p}(s\mid e)$. As an example, the red dashed line in the left panel of Figure \ref{fig:E2T} represents $\E_{\hat{p}(E)}[\widehat{\overline{STE}}(E\ra T)]$, and the three red dashed lines in Figure \ref{fig:S2T} represent $\E_{\hat{p}(S\mid e)}[\widehat{\overline{STE}}(S\ra T\mid e)]$ for $e\in\{-1,0,1\}$. Each figure also displays two boxplots for each measure -- the first shows the distribution of the measure estimated on the bootstrap samples and the second shows the distribution of the measure estimated under the null hypothesis that the causal link in question does not exist (denoted ``H0'').

We begin by considering the total effect of ENSO on temperature shown in Figure \ref{fig:E2T}. Given that $E$ is a root node in the DAG representation given on the right of Figure \ref{fig:climate_dags}, we note that STE and SMI are equivalent, i.e. $STE(e\ra T)=I_1(e;T)$, and the expectation gives an estimate of the mutual information, i.e. $\hat{I}(E;T)=\E_{\hat{p}(E)}[\widehat{STE}(E\ra T)]$. This illustrates the value of considering a specific causal measure -- as we can see, the estimated effect of $E=1$ is roughly three times the effect as estimated by the average with respect to $E$. Recall the interpretation of the SMI as a measure of how much we would expect performing $do(E=e)$ to change the course of nature for $T$. Under this interpretation, we see that forcing an El Ni\~no year would alter the temperature distribution from what we would expect to occur naturally moreso than forcing a La Ni\~na or neutral year.

Figure \ref{fig:E2T} shows that both the direct and indirect effects are less than the STE for all values $e$. This is consistent with the intuition that the direct and indirect effects of ENSO on temperature would not cancel each other out. Intuition is also validated by the fact that the SNIE is less than the SNDE for all values. While this need not be the case in general, we make the assumption that $S$ and $T$ are identically distributed given $E$, and thus we would expect the indirect link $E\ra S \ra T$ to be weaker than the direct link $E\ra T$. While the proposed method does not explicitly identify a physical causal mechanism, the indirect link would represent a situation wherein certain temperatures give rise to environmental circumstances that may affect future temperatures, for example snow pack or soil moisture. Given that there is no evidence in the literature of these environmental factors having a large affect on temperatures, it is sensible that the SNIE is very low. As a final point, we note that while all estimates are statistically significant as measured by our proposed tests, only the effect of $E=1$ has a non-zero lower bound on the confidence interval for the SNIE. This serves as further justification for the measurement of specific causal influences -- when simply measuring average influences with MI, CS, or IF, statistical significance testing results in an ``all or nothing'' test, whereas the present framework enables identifying influences that are significant for only some values of a cause.

We conclude this section with the conditional STE of past on current temperature in a specific ENSO phase, as portrayed by Figure \ref{fig:S2T}. We can clearly see that there is a strong persistence in the temperature anomaly signal, i.e. that the past temperature average has a strong effect on the subsequent average, with the largest effect ($\widehat{\overline{STE}}(S=-1\ra T\mid E=1)$) being roughly five times that of the effect of $E=1$. The fact that the largest effect of $S$ on $T$ occurs when performing the intervention $do(S=-1)$ during an El Ni\~no year can likely be explained by the tendency for El Ni\~no years to give rise to high temperatures. Thus, we would expect that forcing a cold spell during an El Ni\~no would alter the course of nature moreso than, say, forcing a heat wave. Furthermore, the second largest effect is seen when $S=1$ and $E=-1$, i.e. when a heat wave is forced during a La Ni\~na year. This result is reminiscent of the earlier example where there is a large causal influence resulting from a broken chain reaction. In this case, since we would expect an El Ni\~no (resp. La Ni\~na) year to assign a higher probability to a heat wave (resp. cold spell) that would then persist through the effect of $S$ on $T$, intervening on $S$ to force a cold front (resp. heat wave) will result in a large deviation from the natural behavior and thus a large causal effect. It is important to note that ``forcing a cold spell'' is ambiguous in that there are many different mechanisms by which one could hypothetically force a temperature. The following section includes a discussion of how these different mechanisms affect the ability to consider the estimated affect as a true causal effect or merely a measure of predictive utility. In either case, the proposed methods provide a clearer picture of how the relationship between subsequent two-week anomaly averages is modulated by ENSO phase than traditional IT methods. This suggests an area for future investigation, as two-week temperature persistence is not well studied outside of the context of persistent high pressure anomalies \cite{vigaud2018predictability}.

\subsection{Challenges and Caveats}
Any causal interpretation of the results is predicated on the assumption that there are no confounding factors not accounted for in the preprocessing steps. This assumption is less of an issue when measuring the effect of ENSO, where we only need to assume that there is no common cause for $E$ and $S$ or $E$ and $T$ (that there is no backdoor path, to be precise) beyond the seasonality, CO$_2$ forcing, and any other phenomena captured by the leading six harmonics. When measuring the effect of past temperatures, however, this assumption is a bit more far reaching. For example, we have neglected to consider the temperatures in neighboring regions. Moreover, the explicit nature of the causal effect of $S$ on $T$ is more elusive than that of $E$ on $T$. While it is reasonable to expect the temperature to have some causal effect in a literal sense (i.e. via the heat equation), it is likely that the estimation procedure is also capturing the effects of temperature related variables. For example, if we additionally included PNW atmospheric pressure waves in the model, we would expect these waves to be a common cause for $S$ and $T$ resulting in a significantly weaker (if not absent) link $S\ra T$. As such, the above estimate of $\overline{STE}(s\ra T\mid e)$ ought to be viewed as either a measure of predictive utility of the literal temperature, or the causal effect of a ``meta variable'' representative of the temperature and related quantities that are intervened upon as a whole. In any case, the present study serves as a starting point for the development of more intricate causal models relating ENSO and temperature. A potential avenue for continued work is to use the proposed framework on causal graphs learned from data rather than those prespecified using domain expertise. Development of methods for learning causal structures from data is a highly active area of ongoing research in climate science \cite{runge2019inferring}.

A second set of challenges arises from the need to estimate the measures for every value of the cause. While these challenges are indeed a fundamental challenge with the proposed framework, they provide an opportunity for the development of novel estimation and statistical testing techniques. On one hand, the proposed specific causal measures are necessarily more challenging to estimate than their average counterparts. On the other hand, they necessarily provide more resolution and allow for estimating separate confidence intervals for each element in the analysis. If we are trying to estimate $STE(x\ra Y)$ but only have a small number of points in our dataset where $x_n=x$, then we would have very little confidence in our estimate. However, that need not discourage us from having high confidence in an estimate of $STE(x'\ra Y)$ for some $x'$ for which we have many samples. That having been said, the proposed estimators and significance test used in the present study lack a formal analysis and leave considerable room for improvement.

As a final discussion point, we return to the comparison of information theoretic and statistical notions of causal influence. Despite having carefully formulated the proposed measures as measures of the extent to which an intervention results in a deviation from the course of nature, the results presented in this section beg the question: \emph{How useful are bits?} As an absolute measure, it is worth noting that a measure in bits will be largely influenced by the number of quantization regions we select. While this can be partially addressed by the proposed normalization, there is no question that the data compression interpretation provided alongside those equations is less intuitive than a measure of, say, the number of degrees warmer we would expect it to be an El Ni\~no year than a La Ni\~na year. Moreover, this intuition gap would be even larger for someone outside of the information theory community (e.g. climate scientists). This is not to say that the proposed measures are so opaque that they are unusable. In fact, we believe that they provide more interpretable notions of causal influence than other information theoretic measures that have experienced some popularity in the literature. Instead, this discussion is merely to maximize the level of intuition that we can associate with the proposed measures while simultaneously acknowledging the limitations of information theoretic measures in terms of interpretability.

\section{Conclusion}\label{sec:conclusion}
We have sought inspiration from the statistical causality community in order to refine information theoretic measures of causal influence. Specifically, we have developed a series of causal measures that are defined for specific values of the cause in question with the goal of differentiating between total, direct, and indirect effects, and provided conditions under which they can be estimated from observational data. The proposed measures are, at their core, aligned with previous information theoretic measures in that they compare distributions of $Y$ rather than comparing values of $Y$. As such, they are well-equipped for capturing non-linear, higher order causal effects, although at the cost of foregoing an explanation of the exact nature of the causal effects. Perhaps most importantly, we have elucidated the key insight that information theoretic measures of causal influence can be interpreted as methods for quantifying the magnitude with which an intervention is expected to alter the course of nature. This interpretation stands in stark contrast to that of statistical measures. As such, we hope that a key lesson will be that information theoretic and statistical notions of causal can provide \emph{complementary} methods in that they yield the answers to fundamentally different causal questions.
\section*{Acknowledgement}
The CESM project is supported primarily by the National Science Foundation. We thank all the scientists, software engineers, and administrators who contributed to the development of CESM2. All of the data used is made publicly available by NCAR and can be downloaded at \url{https://csegweb.cgd.ucar.edu/exp2-public/cgi-bin/expListPublic.cgi}

G.S. is supported by a postdoctoral fellowship from the Picower Institute for Memory and Learning at the Massachusetts Institute of Technology. S.X. is supported in part by the NSF award AGS-1637450. T.P.C. is supported in part by the Center for Science of Information (CSoI), NSF Science and Technology Center, under grant agreement CCF-0939370; ARO MURI award under contract ARO-W911NF-15-1-0479; NIH award 1R01MH110514; and NSF award IIS-1522125.
\section*{Appendix}
\begin{appendix}

\section{Exchanging Interventions and Observations}\label{app:docalc}
The $do$-calculus provides a set of rules to aid using the $do$-operator in practice and to enable identifying if and how interventional probabilities can be computed. Of particular interest is computing interventional probabilities (i.e. those using the $do$-operator) from the standard conditional probabilities that represent observing variables. This is particularly important in scenarios such as our climate case study, wherein it is infeasible to actually perform interventions. The $do$-calculus consists of three rules, each of which involves an equivalence statement between probabilities that is implied by a d-separation criterion. We here focus on Rule 2, which provides a condition for which observations can be exchanged for actions. Specifically, this rule says that for a DAG $\mc{G}$ and any disjoint sets of variables $X$,$Y$,$Z$, and $W$:
\begin{equation}
(Y {\independent}_d Z \mid X,W)_{\mc{G}_{\overline{X}\underline{Z}}} \implies p(y\mid \hat{x},\hat{z},w)=p(y\mid \hat{x},z,w)
\end{equation}
where $(\cdot {\independent}_d \cdot \mid \cdot)_\mc{G}$ represents d-separation with respect to the DAG $\mc{G}$ and $\mc{G}_{\overline{X}\underline{Z}}$ represents an augmented DAG with all incoming arrows to $X$ and outgoing arrows from $Z$ removed. The rule is framed in a general form in that it allows other variables to be observed or intervened upon (i.e. $W$ and $X$) on both sides of the equality. Roughly speaking, this rule says that if the only way $Z$ relates to $Y$ is via descendants of $Z$, then knowing whether or not a particular value $z$ was observed or forced will not change the distribution of $Y$. To see this, first let $X=\emptyset$, and note that the d-separation condition becomes $(Y {\independent}_d Z \mid W)_{\mc{G}_{\underline{Z}}}$, i.e. $Y$ is d-separated from $Z$ by $W$ if we ignore all paths coming \emph{out} of $Z$. If that condition is not satisfied, then observing a value of $Z$ informs us about the values of $Z$'s parents, which then may provide further information on the distribution of $Y$. By contrast, if we intervene on $Z$, then no information is conveyed about $Z$'s parents, and the distribution of $Y$ will not be the same. Next, letting $X\ne \emptyset$, we see that the condition now requires removing all incoming arrows to $X$. This is because if $X$ is intervened upon, it will contain no information about the values of its parents.

This rule is applied in a straightforward manner in two ways in our case study. First, when measuring the effect of ENSO on temperature, we need to exchange an intervention on the ENSO phase for an observation of an ENSO phase. Focusing on the reduced DAG, the augmented graph $\mc{G}_{\underline{E}}$ is given by $E$ being an isolated node. Thus, in this augmented graph E is d-separated from $T$ by either $\emptyset$ or $S$, and we have $p(t\mid \hat{e})=p(t\mid e)$ and $p(t\mid s,\hat{e})=p(t\mid s,e)$. Similarly, for measuring the effect of $S$ on $T$, we need to consider the augmented graph $\mc{G}_{\underline{S}}$ given by $S\leftarrow E \ra T$. Using the d-separation algorithm described in Algorithm \ref{alg:dsep}, it is straightforward to see that $(S {\independent}_d T \mid E)_{\mc{G}_{\underline{S}}}$ and thus $p(t\mid e,\hat{s})=p(t\mid e,s)$.

\begin{algorithm}[H]
\setlength\baselineskip{15pt}
\caption{d-Separation \cite{lauritzen1990independence}} \label{alg:dsep}
\hspace*{\algorithmicindent} \textbf{Input}: DAG $\mcal{G}=(V,E)$ and disjoint sets $A,B,C\subset V$
\begin{algorithmic}[1]
\State Create a subgraph containing only nodes in $A$, $B$, or $C$ or with a directed path to $A$, $B$, or $C$
\State Connect with an undirected edge any two variables that share a common child
\State For each $c\in C$, remove $c$ and any edge connected to $c$
\State Make every edge an undirected edge
\State Conclude that $A$ and $B$ are d-separated by $C$ if and only if there is no path connecting $A$ and $B$
\end{algorithmic}
\end{algorithm}


\section{Proof of Theorems}\label{app:proofs}

\subsection{Proof of Theorem 1}\label{app:exp_te}
\begin{proof}
The theorem follows directly from the definitions of information flow and STE:
\begin{align*}
\E_{p(X)}[STE(X\ra Y)]
&= \sum_x p(x) \kl{p(Y\mid\hat{x})}{\sum_{x'}p(x')p(Y\mid\hat{x}')} \\
&= \sum_x p(x) \sum_y p(y\mid \hat{x})\log\frac{p(y\mid \hat{x})}{\sum_{x'}p(x')p(y\mid\hat{x}')}\\
&=I(X\ra Y)
\end{align*}
\end{proof}

\subsection{Proof of Theorem 2}\label{app:exp_cde}
\begin{proof}
Starting with the conditional IF, see that:
\begin{align}
I(X\ra Y\mid \hat{Z})
&=\sum_z p(z) \sum_{x}p(x\mid \hat{z})\sum_y p(y\mid \hat{x},\hat{z})\log \frac{p(y\mid \hat{x},\hat{z})}{\sum_{x'}p(x'\mid\hat{z})p(y\mid \hat{x}',\hat{z})}\nonumber\\
&=\sum_{x,z} p(z)p(x\mid \hat{z})\kl{p(y\mid \hat{x},\hat{z})}{\sum\nolimits_{x'}p(x'\mid\hat{z})p(y\mid \hat{x}',\hat{z})}\nonumber\\
&=\sum_{x,z} p(z)p(x)\kl{p(y\mid \hat{x},\hat{z})}{\sum\nolimits_{x'}p(x')p(y\mid \hat{x}',\hat{z})}\label{eq:downstream}\\
&=\E_{p(X)p(Z)}[SCDE(X\ra Y;Z)]\nonumber
\end{align}
where \eqref{eq:downstream} follows from the fact that interventions on $Z$ can be ignored in the distribution of $X$.
Moving onto the CS, we have:
\begin{align*}
\mathfrak{C}_{X\ra Y}
&=\kl{p(X,Y,Z)}{p_{X\ra Y}(X,Y,Z)}\\
&=\sum_{x,y,z}p(x,y,z)\log\frac{p(x)p(z\mid x)p(y\mid x,z)}{p(x)p(z\mid x)\left(\sum_{x'}p(x')p(y\mid x',z)\right)}\\
&=\sum_{x,y,z}p(x,y,z)\log\frac{p(y\mid x,z)}{\sum_{x'}p(x')p(y\mid x',z)}\\
&=\sum_{x,z}p(x,z)\sum_y p(y\mid x,z)\log\frac{p(y\mid x,z)}{\sum_{x'}p(x')p(y\mid x',z)}\\
&=\sum_{x,z}p(x,z)\sum_y p(y\mid \hat{x},\hat{z})\log\frac{p(y\mid \hat{x},\hat{z})}{\sum_{\hat{x}'}p(\hat{x}')p(y\mid \hat{x}',\hat{z})}\\
&=\sum_{x,z}p(x,z)\kl{p(Y\mid \hat{x},\hat{z})}{\sum\nolimits_{x'}p(x')p(Y\mid \hat{x}',\hat{z})}\\
&=\E_{p(X,Z)}[SCDE(X\ra Y;Z)] 
\end{align*}
\end{proof}

\subsection{Proof of Theorem 3}\label{app:idendifiability}
\begin{proof}
Note that the conditional STE, SNDE, and SNIE only utilize three distributions involving interventions, namely $p(y\mid \hat{x},\tilde{u})$, $p(z\mid \hat{x},\tilde{u})$, and $p(y\mid \hat{x},z,\tilde{u})$. We wish to show that we can estimate these distributions can be estimated from observational data, i.e. that the hats can be removed. Assume that the conditions of the theorem hold. We first claim that $(X \independent Y \mid \tilde{U}_1)_{\mcal{G}_{\underline{X}}} \implies (X \independent Y \mid \tilde{U})_{\mcal{G}_{\underline{X}}}$ and $(X \independent Z \mid \tilde{U}_2)_{\mcal{G}_{\underline{X}}} \implies (X \independent Z \mid \tilde{U})_{\mcal{G}_{\underline{X}}}$. To see this, note that in the DAG $\mc{G}_{\underline{X}}$, $X$ has no children, and thus will not be connected to any other nodes in step two of the d-separation algorithm given by Algorithm \ref{alg:dsep}. Since every edge connected to a node in $\tilde{U}$ is removed in step three in the algorithm, the only way for one of the implications to be violated is if there is an undirected path in $\mc{G}_{\underline{X}}$ connecting $X$ and $Z$ or $X$ and $Y$ that does not pass through $\tilde{U}$; however, such a path would necessarily not pass through $\tilde{U}_1$ or $\tilde{U}_2$, which would violate $(X \independent Y \mid \tilde{U}_1)_{\mcal{G}_{\underline{X}}}$ or $(X \independent Z \mid \tilde{U}_2)_{\mcal{G}_{\underline{X}}}$. Thus, the claimed implications hold. Next we can directly apply rule two of the $do$-calculus \cite[Theorem 3.4.1]{pearl2009causality} to $(X \independent Y \mid \tilde{U})_{\mcal{G}_{\underline{X}}}$ and $(X \independent Z \mid \tilde{U})_{\mcal{G}_{\underline{X}}}$ to see that $p(y\mid \hat{x},\tilde{u})=p(y\mid x,\tilde{u})$ and $p(z\mid \hat{x},\tilde{u})=p(z\mid x,\tilde{u})$. Finally, we claim that $(X \independent Y \mid \tilde{U})_{\mcal{G}_{\underline{X}}}\implies (X \independent Y \mid Z,\tilde{U})_{\mcal{G}_{\underline{X}}}$ using the same argument showing the implications above. Applying rule 2 of the $do$-calculus to $(X \independent Y \mid Z,\tilde{U})_{\mcal{G}_{\underline{X}}}$ yields that $p(y\mid \hat{x},z,\tilde{u})=p(y\mid x,z,\tilde{u})$. As such, all three of the interventional distributions needed by the STE, SNDE, and SNIE can be equated to their observational counterparts under the stated assumptions and the proof is completed. 
\end{proof}


\section{Conditional Specific Causal Measures}\label{app:settingspecific}

\begin{Definition}
The partially observed conditional SCDE of $x$ on $Y$ with mediator $z$ given $\tilde{u}$ is defined as:
\begin{equation*}
SCDE(x\ra Y;z \mid \tilde{u})
\triangleq \kl{p(Y\mid \hat{x},\hat{z},\tilde{u})}{\sum\nolimits_{x'} p(x'\mid \tilde{u})p(Y\mid \hat{x}',\hat{z},\tilde{u})}
\end{equation*}
\noindent In the fully observable setting $\tilde{U}=U$ we have:
\begin{equation*}
SCDE(x\ra Y;z\mid u)\triangleq \kl{p(Y\mid \hat{x},\hat{z},u_Y)}{\sum\nolimits_{x'} p(x'\mid u_X)p(Y\mid \hat{x}',\hat{z},u_Y)}
\end{equation*}
\end{Definition}

\begin{Definition}
The partially observed conditional SNDE of $x$ on $Y$ given $\tilde{u}$ is defined as:
\begin{equation*}
SNDE(x\ra Y\mid \tilde{u})
\triangleq \kl{p(Y\mid \hat{x},\tilde{u})}
{\sum\nolimits_{x',z'} p(x'\mid \tilde{u})p(z'\mid \hat{x},\tilde{u})p(Y\mid \hat{x}',z',\tilde{u})}
\end{equation*}
\noindent In the fully observable setting $\tilde{U}=U$ we have:
\begin{equation*}\label{eq:SSNDE}
SNDE(x\ra Y\mid u)
\triangleq \kl{p(Y\mid \hat{x},u_Y,u_Z)}
{\sum\nolimits_{x',z'} p(x'\mid u_X)p(z'\mid \hat{x},u_Z)p(Y\mid \hat{x}',z',u_Y)}
\end{equation*}
\end{Definition}

\begin{Definition}
The partially observed conditional SNIE of $x$ on $Y$ given $\tilde{u}$ is defined as:
\begin{equation*}
SNIE(x\ra Y\mid \tilde{u})
\triangleq\kl{p(Y\mid \hat{x},\tilde{u})}
{\sum\nolimits_{x',z'} p(x'\mid \tilde{u})p(z'\mid \hat{x}',\tilde{u})p(Y\mid \hat{x},z',\tilde{u})}
\end{equation*}\noindent In the fully observable setting $\tilde{U}=U$ we have:
\begin{equation*}
SNIE(x\ra Y\mid u)
\triangleq\kl{p(Y\mid \hat{x},u_Y,u_Z)}
{\sum\nolimits_{x',z'}p(x'\mid u_X)p(z' \mid \hat{x}',u_Z)p(Y\mid \hat{x},z',u_Y)}
\end{equation*}
\end{Definition}


\section{Normalized Specific Causal Measures}\label{app:normalization}

\begin{Definition}
The normalized conditional SCDE of $x$ on $Y$ given $\tilde{u}$ is defined as:
\begin{equation*}
\overline{SCDE}(x\ra Y\mid \tilde{u};z) \triangleq \frac{SCDE(x\ra Y\mid \tilde{u};z)}{SCDE(x\ra Y\mid \tilde{u};z)+H(Y \mid do(X=x),do(Z=z),\tilde{U}=\tilde{u})}
\end{equation*}
\end{Definition}

\begin{Definition}
The normalized conditional SNDE of $x$ on $Y$ given $\tilde{u}$ is defined as:
\begin{equation*}
\overline{SNDE}(x\ra Y\mid \tilde{u}) \triangleq \frac{SNDE(x\ra Y\mid \tilde{u})}{SNDE(x\ra Y\mid \tilde{u})+H(Y \mid do(X=x),\tilde{U}=\tilde{u})}
\end{equation*}
\end{Definition}

\begin{Definition}
The normalized conditional SNIE of $x$ on $Y$ given $\tilde{u}$ is defined as:
\begin{equation*}
\overline{SNIE}(x\ra Y\mid \tilde{u}) \triangleq \frac{SNIE(x\ra Y\mid \tilde{u})}{SNIE(x\ra Y\mid \tilde{u})+H(Y \mid do(X=x),\tilde{U}=\tilde{u})}
\end{equation*}
\end{Definition}

\section{Additional Details on Maximum Likelihood Estimation}\label{app:estimation}

For an arbitrary collection of $N$ samples $\mc{C}=(x_n,y_n,z_n)_{n=1}^N$ of variables $X,Y,Z$, the maximum likelihood estimate of the (conditional) pmf of $Y$ (given $x$ and/or $z$) is given by:
\begin{gather*}
\hat{p}_\mc{C}(y)\triangleq \frac{\left|\{n:y_n=y\}\right|}{n} \ \ \ \ \ \ \ \ \ \ \ \
\hat{p}_\mc{C}(y \mid x)\triangleq \frac{\left|\{n:x_n=x,y_n=y\}\right|}{\left|\{n:x_n=x\}\right|} \label{eq:mle_cond}\\
\hat{p}_\mc{C}(y \mid x,z)\triangleq \frac{\left|\{n:x_n=x,y_n=y,z_n=z\}\right|}{\left|\{n:x_n=x,z_n=z\}\right|}
\end{gather*}
\noindent where the $\left|\{\cdot\}\right|$ gives the number of elements in the set $\{\cdot\}$.

The normalized estimates are given by:
\begin{gather*}
\widehat{\overline{STE}}_\mc{D}(e\ra T)\triangleq\frac{\widehat{STE}_\mc{D}(e\ra T)}{\widehat{STE}_\mc{D}(e\ra T)+\hat{H}_\mc{D}(T\mid e)} \\
\widehat{\overline{SNDE}}_\mc{D}(e\ra T)\triangleq\frac{\widehat{SNDE}_\mc{D}(e\ra T)}{\widehat{SNDE}_\mc{D}(e\ra T)+\hat{H}_\mc{D}(T\mid e)}\\
\widehat{\overline{SNIE}}_\mc{D}(e\ra T)\triangleq\frac{\widehat{SNIE}_\mc{D}(e\ra T)}{\widehat{SNIE}_\mc{D}(e\ra T)+\hat{H}_\mc{D}(T\mid e)} \\
\widehat{\overline{STE}}_\mc{D}(s\ra T\mid e)\triangleq\frac{\widehat{STE}_\mc{D}(s\ra T\mid e)}{\widehat{STE}_\mc{D}(s\ra T\mid e)+\hat{H}_\mc{D}(T\mid e,s)}
\end{gather*}
where $\hat{H}_\mc{D}(T\mid e)\triangleq -\sum_t\hat{p}_\mc{D}(t\mid e)\log \hat{p}_\mc{D}(t\mid e)$ and $\hat{H}_\mc{D}(T\mid e,s)\triangleq -\sum_t\hat{p}_\mc{D}(t\mid e,s)\log \hat{p}_\mc{D}(t\mid e,s)$. In all of the figures, we have multiplied the estimated measures by 100 to obtain a percentage.


\section{Climate Model Details}
In concordance with the CMIP6 terms of use (\url{https://pcmdi.llnl.gov/CMIP6/TermsOfUse/TermsOfUse6-1.html}), we provide the full model details for the model that provided the utilized dataset.
\begin{center}
\begin{longtable}{ | m{6cm} | m{6cm} | }
 \hline
 source id & CESM2  \\
 \hline
 institution id & NCAR \\
 \hline
 release year & 2018 \\
 \hline
 activity participation & AerChemMIP C4MIP CDRMIP CFMIP CMIP CORDEX DAMIP DCPP DynVarMIP GMMIP GeoMIP HighResMIP ISMIP6 LS3MIP LUMIP OMIP PAMIP PMIP RFMIP SIMIP ScenarioMIP VIACSAB VolMIP \\
 \hline
 cohort & Registered \\
 \hline
 label & CESM2 \\
 \hline
 label extended & CESM2 \\
 \hline
 atmos & CAM6 (0.9x1.25 finite volume grid; 288 x 192 longitude/latitude; 32 levels; top level 2.25 mb) \\
 \hline
 natNomRes atmos & 100 km \\
 \hline
 ocean & POP2 (320x384 longitude/latitude; 60 levels; top grid cell 0-10 m) \\
 \hline
 natNomRes ocear & 100 km \\
 \hline
 landIce & CISM2.1 \\
 \hline
 natNomRes landIce & 5 km \\
 \hline
 aerosol & MAM4 (same grid as atmos) \\
 \hline
 atmosChem & MAM4 (same grid as atmos) \\
 \hline
 land & CLM5 (same grid as atmos) \\
 \hline
 ocnBgchem & MARBL (same grid as atmos) \\
 \hline
 seaIce & CICE5.1 (same grid as atmos) \\
 \hline
\end{longtable}
\end{center}
\end{appendix}

\bibliography{references.bib}

\end{document}